\documentclass[aps,prl,twocolumn,reprint,amsmath,amssymb,superscriptaddress,floatfix,footinbib,longbibliography]{revtex4-2}

\usepackage{graphicx}
\usepackage{epstopdf}
\usepackage{afterpage}
\usepackage[T1]{fontenc}
\usepackage[applemac]{inputenc}
\usepackage{lmodern}
\usepackage{amsmath}
\usepackage{amssymb}
\usepackage{natbib}
\usepackage{ae}
\usepackage{units}
\usepackage{upgreek}

\usepackage{amsmath,amssymb,natbib,bm}
\usepackage{psfrag}
\usepackage{subfigure}
\usepackage{amsthm}

\usepackage{booktabs}

\usepackage{slashed}

\usepackage[americaninductors]{circuitikz}
\usepackage{tikz}
\usetikzlibrary{arrows}
\usepackage{ulem}
\usepackage{color}
\usepackage{url}

\usepackage[colorlinks]{hyperref}
\hypersetup{%
        plainpages=true,
        breaklinks=true,
        hypertexnames=false,
        pageanchor=true,
        colorlinks=true,
        linkcolor={blue},
        citecolor={magenta},
        urlcolor={blue},
        anchorcolor={black}
      }

\usepackage{mleftright} 

\renewcommand{\eqref}[1]{\mbox{Eq.~(\ref{#1})}}

\newcommand{\figpanel}[2]{Fig.~\hyperref[#1]{\ref*{#1}(#2)}} 
\newcommand{\figpanels}[3]{Figs.~\hyperref[#1]{\ref*{#1}(#2)-(#3)}} 
\newcommand{\figpanelNoPrefix}[2]{\hyperref[#1]{\ref*{#1}(#2)}} 
\newcommand{\figpanelsNoPrefix}[3]{\hyperref[#1]{\ref*{#1}(#2)-(#3)}} 

\newcommand{\figand}[3]{Figs.~\hyperref[#1]{\ref*{#1}(#2)} and \hyperref[#1]{\ref*{#1}(#3)}}

\newcommand{\be}{\begin{equation}}
\newcommand{\ee}{\end{equation}}
\newcommand{\bea}{\begin{eqnarray}}
\newcommand{\eea}{\end{eqnarray}}

\begin{document}
\title{Collective Floquet engineering between two distant superconducting artificial atoms}

\author{Y.-H.~Huang}
\thanks{These authors contributed equally}
\affiliation{Department of Physics, National Tsing Hua University, Hsinchu 30013, Taiwan}

\author{Jiale~Yuan}
\thanks{These authors contributed equally}
\email[e-mail:]{jialeyuan@csu.edu.cn}
\affiliation{School of Physics, Central South University, Changsha 410083, China}
\affiliation{Zhejiang Key Laboratory of Micro-Nano Quantum Chips and Quantum Control, School of Physics, and State Key Laboratory for Extreme Photonics and Instrumentation, Zhejiang University, Hangzhou 310027, China}

\author{Z.~Q.~Niu}
\thanks{These authors contributed equally}
\affiliation{State Key Laboratory of Materials for Integrated Circuits, Shanghai Institute of Microsystem and Information Technology, Chinese Academy of Sciences, Shanghai 200050, China}

\author{Y.-S.~Tsai}
\affiliation{Department of Physics, National Tsing Hua University, Hsinchu 30013, Taiwan}

\author{K.-C.~Yu}
\affiliation{Department of Physics, National Tsing Hua University, Hsinchu 30013, Taiwan}

\author{P.-H.~Hsu}
\affiliation{Department of Physics, National Tsing Hua University, Hsinchu 30013, Taiwan}

\author{K.-M.~Hsieh}
\affiliation{Department of Physics, City University of Hong Kong, Kowloon, Hong Kong SAR 999077, China}

\author{Haw-Tyng~Huang}
\affiliation{Department of Materials Science and Engineering, National Taiwan University, Taipei, 10617, Taiwan}

\author{J.~C.~Chen}
\affiliation{Department of Physics, National Tsing Hua University, Hsinchu 30013, Taiwan}
\affiliation{Center for Quantum Technology, National Tsing Hua University, Hsinchu 30013, Taiwan}

\author{P.~Y.~Wen}
\affiliation{Department of Physics, National Chung Cheng University, Chiayi 621301, Taiwan}

\author{Y.-H.~Lin}
\affiliation{Department of Physics, National Tsing Hua University, Hsinchu 30013, Taiwan}
\affiliation{Center for Quantum Technology, National Tsing Hua University, Hsinchu 30013, Taiwan}

\author{Z.-R.~Lin}
\email[e-mail:]{zrlin@mail.sim.ac.cn}
\affiliation{State Key Laboratory of Materials for Integrated Circuits, Shanghai Institute of Microsystem and Information Technology, Chinese Academy of Sciences, Shanghai 200050, China}

\author{Da-Wei~Wang}
\affiliation{Zhejiang Key Laboratory of Micro-Nano Quantum Chips and Quantum Control, School of Physics, and State Key Laboratory for Extreme Photonics and Instrumentation, Zhejiang University, Hangzhou 310027, China}

\author{I.-C.~Hoi}
\email[e-mail:]{iochoi@cityu.edu.hk}
\affiliation{Department of Physics, City University of Hong Kong, Kowloon, Hong Kong SAR 999077, China}

\date{\today}

\begin{abstract}
A one-dimensional waveguide is a crucial medium for transferring propagating quantum information between remote qubits, yet the photon-mediated interaction is typically fixed by the static device geometry, qubit parameters, and their couplings to the transmission line. Such an interaction can be engineered to be coherent or dissipative via virtual- or real-photon exchange, respectively, yet its effective strength remains fixed in either case. 
Here we demonstrate dynamical control over this interaction by applying Floquet engineering to two superconducting qubits coupled to a waveguide terminated by a capacitive mirror. By positioning one qubit at an antinode and the other qubit at a node or an antinode of the resonant standing wave, we realize a Hermitian or non-Hermitian coupling between the two qubits. A periodic flux drive of the qubit frequencies then enables on-demand control of the coherent energy splitting and the collective decay. Our results establish Floquet engineering as a route to dynamically reshape both coherent and dissipative photon-mediated interactions between remote quantum emitters.


\end{abstract}
\maketitle


\afterpage{%
\begin{figure*}[t]
\includegraphics[width=\linewidth]{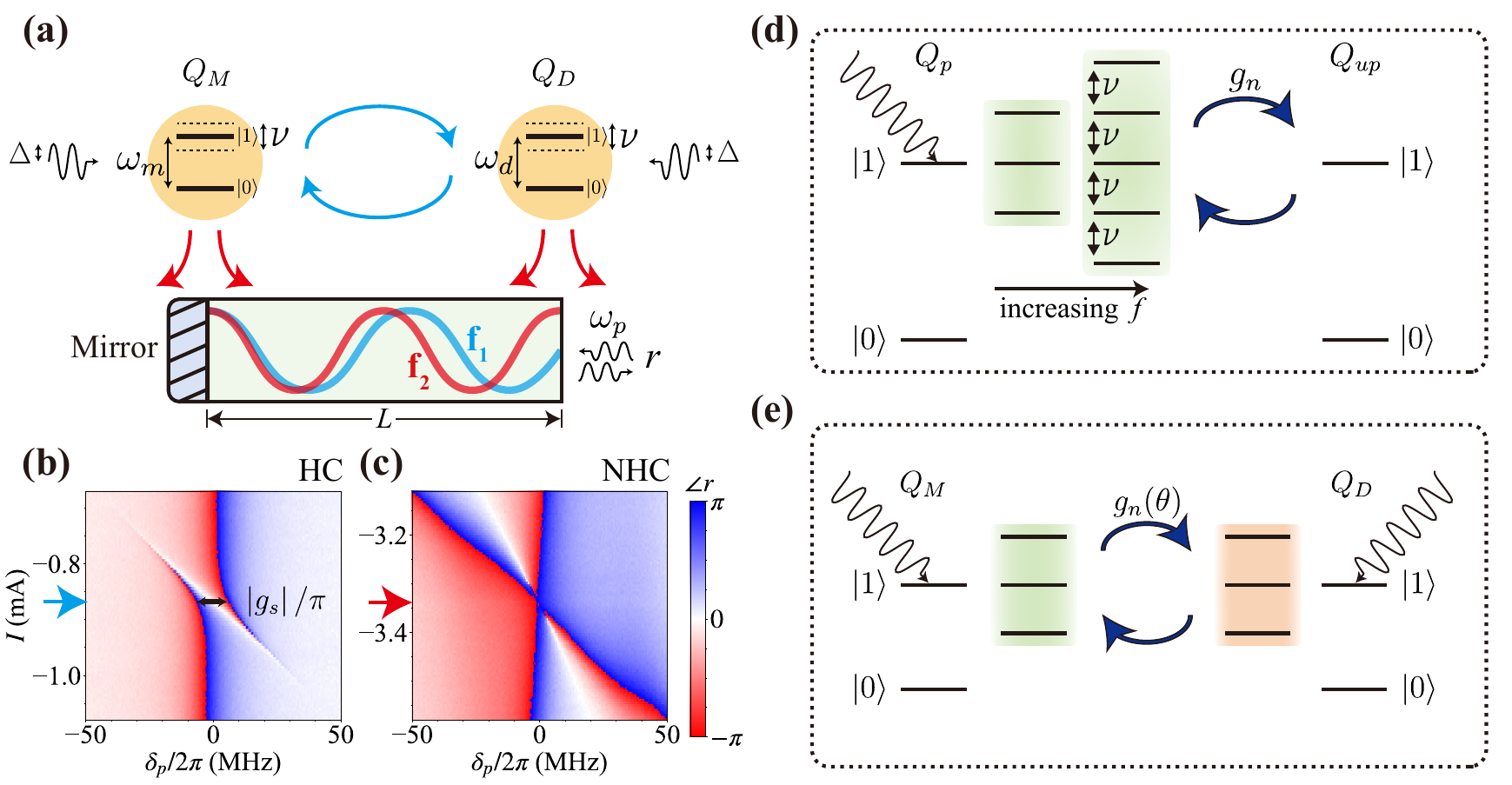}
\caption{System schematic, static interaction, and Floquet level diagrams. (a) Two transmon qubits $Q_M$ and $Q_D$ with resonant transition frequency $\omega_{m,d}$ are capacitively coupled to a one-dimensional semi-infinite transmission line terminated by a mirror. $Q_M$ is fixed at an antinode near the mirror and is maximally coupled to the transmission line. $Q_D$ can be effectively placed at a node or an antinode by tuning its transition frequency. The device layout and experimental setup are shown in Fig.~S1, while the individual qubit characterization is provided in Fig.~S2 and Table~S1 of the Supplemental Material \cite{SupMat}.
At $f_1/2\pi = 4.527$ GHz (blue wave), $Q_D$ sits at a node, resulting in HC; at $f_2/2\pi = 5.031$ GHz (red wave), $Q_D$ sits at an antinode, resulting in NHC. The system is probed by a weak continuous-wave field at frequency $\omega_p$ incident from the open end, and the reflection coefficient $r$ is measured. (b)-(c) Phase response $\angle r$ measured as a function of probe detuning $\delta_p$ and the bias current $I$ applied to $Q_D$, with $\omega_m$ fixed at $f_1$ in (b) and $f_2$ in (c) as $\omega_d$ is tuned through resonance.
In (b), an avoided level crossing is observed, indicating the HC configuration.
In (c), a level crossing appears with an enhanced
collective linewidth, indicating the NHC configuration. The corresponding magnitude responses $\left |r\right|$ are shown in Fig.~S3 of the Supplemental Material \cite{SupMat}.
(d) Floquet level diagrams for a single driven qubit. $Q_p$ and $Q_{up}$ denote the pumped and unpumped qubits, respectively. The pumped qubit ($Q_p$) splits into Floquet states separated by energy $\nu$ and the $n$-th Floquet state couples to the unpumped qubit ($Q_{up}$) with strength $g_n$ either coherently or dissipatively, depending on the operating frequency $f_1$ or $f_2$. 
(e) Floquet level diagram for the case when both qubits are driven. The Floquet states of the two qubits couple either coherently or dissipatively, with coupling strengths governed by the relative pump phase $\theta$.
} 
\label{fig:1}
\end{figure*}
}

\textit{Introduction}---The Lamb shift, a small energy shift between the $2S_{1/2}$ and $2P_{1/2}$ levels of hydrogen \cite{lamb1947fine}, provided one of the earliest and most striking signatures of quantum vacuum fluctuations.  An isolated atom constantly emits and reabsorbs virtual photons, thereby shifting its energy levels.  When two or more atoms share a common electromagnetic environment, they can exchange virtual photons  \cite{van2013photon}, giving rise to a collective Lamb shift (CLS) \cite{wen2019large}, which is a coherent frequency shift that directly reveals vacuum-mediated interactions between distinct emitters. In contrast, the exchange of real photons induces collective radiative phenomena such as Dicke superradiance and subradiance \cite{dicke1954coherence, mlynek2014observation, zanner2022coherent}. These two configurations, involving virtual- and real-photon exchange, 
can be tuned by controlling the positions of the emitters and the structure of the vacuum modes \cite{van2013photon, bao2024level, sheremet2023waveguide}.

A notable advance in CLS came from a waveguide quantum electrodynamics (QED) setup in which two superconducting qubits were positioned in front of a mirror \cite{wen2019large}. By positioning a distant qubit at a node of the resonant standing wave, its radiative relaxation into the waveguide was suppressed, while the virtual-photon exchange with the qubit at the mirror persisted. This enabled a clean, static measurement of the CLS, achieving a shift twice the transition linewidth. The same mirror-waveguide geometry also provides access to the complementary configuration mediated by real photons, where both qubits are positioned at antinodes of the standing wave and their emissions become correlated, producing superradiant and subradiant collective decay channels. However, in these static waveguide-QED configurations, the strength and character of the photon-mediated interaction are determined by the qubits' transition frequencies, their positions within the electromagnetic mode structure, and the resulting couplings to the transmission line \cite{wen2019large, van2013photon, lalumiere2013input, lin2019scalable, hoi2015probing, run2026realizing}. Achieving dynamic control over both virtual-photon- and real-photon-mediated collective interactions remains an open challenge.

Floquet engineering generates effective Hamiltonians unattainable in static systems \cite{goldman2014periodically}. This approach has been applied across diverse quantum platforms. In cold atoms, this technique creates artificial gauge fields and topological bands \cite{jotzu2014experimental, wang2024floquet, an2017direct, xu2022floquet, wang2015topological}. In photonic systems and two-dimensional materials, it produces Floquet topological insulators and light-induced Chern phases \cite{rechtsman2013photonic, mciver2020light, maczewsky2017observation, afzal2020realization, oka2009photovoltaic, deng2022}. 
 In hybrid magnonic systems, Floquet engineering enables coherent control of magnon-microwave-photon coupling \cite{xu2020floquet}.
 More broadly, it enables the synthesis of non-equilibrium many-body states, including discrete time crystals \cite{zhang2017observation,zhang2022digital}, dynamical topological phases \cite{dumitrescu2022dynamical}, and engineered vibrational dynamics  \cite{kiefer2019floquet}. Crucially, a periodic drive restructures the Hamiltonian \cite{goldman2014periodically}, offering a versatile tool to control interactions \cite{nguyen2024programmable, liu2020synthesizing}, engineer band structures \cite{rechtsman2013photonic, jotzu2014experimental, rudner2020band, wang2025realizing}, and even tailor dissipation on demand \cite{petiziol2022cavity, harrington2022engineered}.


In this Letter, we demonstrate Floquet engineering of the complex photon-mediated interaction between two distant superconducting artificial atoms in a mirror-terminated waveguide. Superconducting waveguide-QED platforms combine strong light-matter coupling with in situ tunability of artificial-atom transition frequencies, and have enabled a broad range of quantum-optical phenomena, including deterministic photon loading \cite{lin2022deterministic, cheng2024tuning}, Landau-Zener-St\"{u}ckelberg-Majorana interferometry \cite{wen2020landau, liul2023coherent}, and controllable group delay \cite{cheng2025group}. By choosing the operating frequency, we can access two configurations: Hermitian coupling (HC) via virtual photons and non-Hermitian coupling (NHC) via real photons. A periodic magnetic flux drive of the qubit frequency enables dynamical control of both the CLS and the radiative decay. We probe these driven interactions through the reflection spectrum of a weak coherent field \cite{lalumiere2013input}. 
Moreover, when both qubits are driven, the relative phase between the two drives serves as an independent control parameter that continuously tunes the qubit-qubit interaction without altering the qubit frequencies, their spatial separation, or the pump power.
These tunabilities allow us to suppress, restore, and reshape remote photon-mediated interactions, providing a route toward on-demand control of distant qubits, with potential relevance to quantum state transfer and remote entanglement in distributed quantum architectures \cite{bienfait2019phonon, zhang2022synthesizing, song2025realization, saha2025high}.

\afterpage{%
\begin{figure*}[t]
\includegraphics[width=\linewidth]{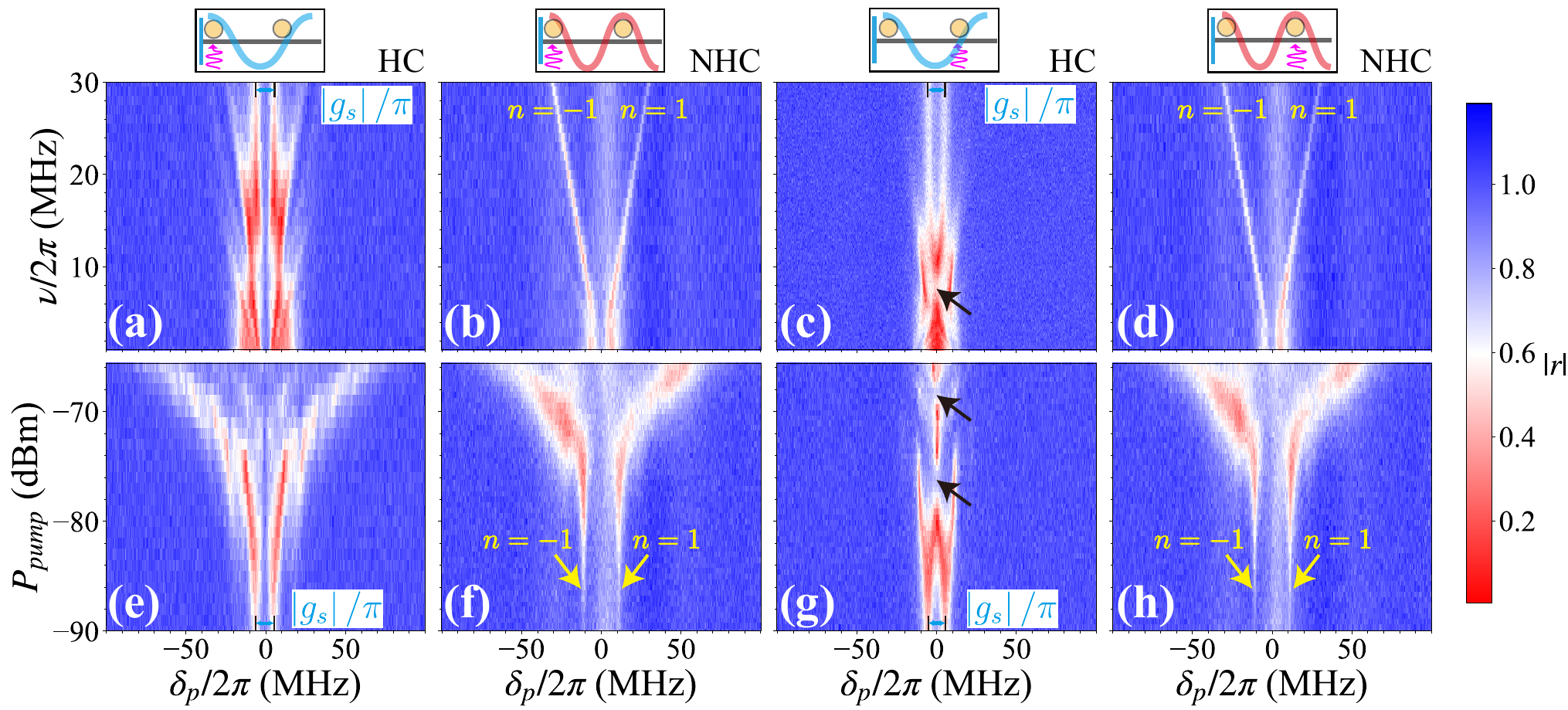}
\caption{Floquet-engineered qubit-qubit coupling with a single driven qubit. The coupling mechanism is illustrated in \figpanel{fig:1}{d}. Reflection amplitude $\left | r\right |$ is plotted as a function of probe detuning $\delta_p$ and either modulation frequency $\nu$ (upper panels) or pump power $P_{pump}$ (lower panels). 
The upper insets show the coupling and the drive configurations.
In panels (a), (b), (e), and (f), $Q_M$ is pumped; in panels (c), (d), (g), and (h), $Q_D$ is pumped. Data in (a), (c), (e), and (g) are measured at $f_1$, the node frequency of $Q_D$, where the Floquet bands split due to the virtual-photon exchange in the HC configuration, as is also evident from the calculated eigenvalue branches in Figs.~S4(e), S4(g), S7(e), and S7(g) of the Supplemental Material \cite{SupMat}. Data in (b), (d), (f), and (h) are measured at $f_2$, the antinode frequency of $Q_D$, corresponding to the NHC configuration, where the linewidths of the Floquet bands are associated with the correlated decay rates. Corresponding numerical results are shown in Figs.~S4(b), S4(d), S7(b), S7(d) of the Supplemental Material \cite{SupMat}. 
In (c) and (g), the arrows mark the collapse of the zeroth-band splitting near the zeros of $J_0(f)$, indicating coherent destruction of tunneling. The upper panels (a)-(d) are measured at a fixed pump power $P_{pump}\approx\unit[-80]{dBm}$ and the lower panels (e)-(h) are measured at a fixed modulation frequency $\nu/2\pi=\unit[11]{MHz}$. 
}
\label{fig:2}
\end{figure*}
}

\textit{Experimental setup and static response}---Our waveguide QED system is depicted schematically in \figpanel{fig:1}{a}. 
Two qubits are coupled through a one-dimensional semi-infinite transmission line that is capacitively coupled to ground at one end, effectively forming a mirror termination for the waveguide. The mirror gives rise to a standing wave in the transmission line  \cite{hoi2015probing}, allowing us to control the qubit-qubit coupling by tuning the qubits' transition frequencies \cite{wen2019large, lin2019scalable}. In our device, one qubit is placed at the mirror and is referred to as the mirror qubit ($Q_M$). It always sits at an antinode of the standing wave and is therefore maximally coupled to the waveguide. The other qubit, located approximately  $L=60$ {mm} away from the mirror, remains within the Markovian regime ($2L/v  \ll \Gamma_{10}^{-1}$) and is referred to as the distant qubit ($Q_D$). Although the position of $Q_D$ is fixed, we can vary its transition frequency $\omega_{d}$ to adjust its normalized location $L/\lambda$, where $\lambda=2\pi v/\omega_{d}$ is the wavelength and $v \sim \unit[1.23 \times 10^{8}]{m/s}$ is the speed of light in the waveguide.  This allows us to effectively position $Q_D$ at either a node or an antinode, yielding markedly distinct physical configurations. 

Tracing out the vacuum modes of the waveguide yields an effective Hamiltonian for the qubits
\begin{equation}
H = \frac{\omega_d}{2}\sigma_d^z + \frac{\omega_m}{2}\sigma_m^z + g_s (\sigma_m^+\sigma_d^- + \sigma_d^+\sigma_m^-),
\end{equation}
where $\omega_m$ and $\omega_d$ are the transition frequencies of $Q_M$ and $Q_D$, $g_s$ is the effective qubit-qubit coupling strength, and $\sigma^z_{m,d}$, $\sigma^+_{m,d}$, and $\sigma^-_{m,d}$ denote the $z$-component of the Pauli matrices, the Pauli raising operator, and the Pauli lowering operator, respectively.
When both qubits are tuned to $f_1/2\pi = \unit[4.527]{GHz}$, $Q_D$ is at a node, as indicated by the blue wave in \figpanel{fig:1}{a}. The coupling $g_s = \sqrt{\Gamma_m(f_1)\Gamma_d(f_1)}$ is then real, with $\Gamma_{m,d}(\omega)$ being the bare relaxation rates \cite{lin2019scalable}. This HC configuration supports a CLS. We probe the system by measuring the reflection coefficient $r$ for a weak probe field, whose Rabi frequency $\Omega$ is much smaller than the qubit decoherence rate, such that the population induced by the probe field is negligible. The resulting reflection spectrum [\figpanel{fig:1}{b}] displays an avoided level crossing with a splitting of $2g_s$, directly revealing a CLS comparable to twice the transition linewidth \cite{wen2019large}. Here the inter-qubit coupling is mediated by off-resonant virtual modes, since $Q_D$, sitting at a node, decouples from the resonant field.
When both qubits are instead tuned to $f_2/2\pi = \unit[5.031]{GHz}$, $Q_D$ sits at an antinode, as indicated by the red wave in \figpanel{fig:1}{a}. In this case $g_s = -i\sqrt{\Gamma_m(f_2)\Gamma_d(f_2)}$ is purely imaginary, making the coupling non-Hermitian. The qubits now interact through the exchange of real photons, and the spectrum exhibits a level crossing with an enhanced
collective linewidth [\figpanel{fig:1}{c}], consistent with a dissipative coupling configuration mediated by real photons \cite{lalumiere2013input, bao2024level}. Detailed qubit parameters are provided in Table S1 of the Supplemental Material \cite{SupMat}.

\textit{Floquet engineering}---Static waveguide-QED geometries restrict photon-mediated interactions to fixed values set by the qubits' transition frequencies and the device geometry. Floquet engineering lifts this constraint by introducing dynamical control through periodic modulation of the qubit energies. In our experiment, this unlocks real-time, in situ tuning of both the coherent CLS and the dissipative collective decay rate, while revealing quantum-interference phenomena inaccessible in the undriven system. 

We apply a sinusoidal flux pump to each qubit, so that its transition frequency is modulated as $\omega_i(t) = f_j + \Delta_i \cos(\nu t + \theta_i)$, where $\nu$ is the pump frequency, $\Delta_i$ the amplitude, $\theta_i$ the phase, and $j=1,2$ selects the node (HC) or antinode (NHC) working point. Under this drive, the energy levels of a qubit split into Floquet bands spaced by $\nu$ [\figand{fig:1}{d}{e}]. Tracing out the waveguide modes yields effective coupling strengths between the Floquet states of the two qubits, governed by the modulation parameters. Importantly, these Floquet sidebands do not merely replicate the driven-qubit spectrum; the periodic modulation directly renormalizes the photon-mediated inter-qubit interaction, allowing its coherent and dissipative components to be dynamically reshaped.

When only one qubit is driven [\figpanel{fig:1}{d}], the coupling between the $n$-th Floquet band of the pumped qubit and the unpumped qubit takes the form (Supplemental Material \cite{SupMat}, Sec.~S4B) \cite{li2013motional}
\begin{equation}
g_n = g_s\,J_n(f),\qquad f = \Delta_i/\nu,
\label{eq:onedrive}
\end{equation}
where $J_n$ is the $n$-th order Bessel function of the first kind. The Bessel factor therefore directly controls the amplitude of each Floquet-mediated coupling channel rather than merely shifting the qubit spectrum. By simply varying the pump frequency $\nu$ or amplitude $\Delta_i$, we can tune $g_n$ from its static value down to zero. Figures~\figpanelsNoPrefix{fig:2}{a}{d} showcase this tunability through the reflection amplitude versus probe detuning $\delta_p$ and $\nu$ at fixed pump power. At the node frequency of $Q_D$, i.e., in the HC configuration [\figand{fig:2}{a}{c}], the Floquet sidebands appear at \(\delta_p = n\nu\), with splittings governed by the Bessel-dressed coupling, as further illustrated by the calculated eigenvalue curves in Figs.~S4(e) and S4(g) of the Supplemental Material \cite{SupMat}. In the small-\(f\) limit, the band splittings are approximately $2 \left |g_s \right |$. Notably, the collapse of the band splitting near the zeros of $J_0(f)$ [black arrows in \figpanel{fig:2}{c}] demonstrates coherent destruction of tunneling (CDT) \cite{grossmann1991coherent, ashhab2007two} in a waveguide-QED context. The physical origin of CDT is the destructive interference of multiple Floquet-mediated exchange pathways. Consequently, virtual-photon exchange, and hence the CLS, can be dynamically switched off, even though the qubits remain coupled to the same waveguide continuum. This provides a mechanism for dynamically isolating remote qubits without physically detuning or repositioning them, offering a potential route to suppressing idle crosstalk in distributed or modular quantum architectures. 

At the antinode frequency of $Q_D$, the system operates in the NHC configuration [\figand{fig:2}{b}{d}], where the coupling is purely dissipative. Although a weak stray resonance [see Fig.~S3(b) of the Supplemental Material \cite{SupMat}] obscures the direct linewidth signal in the data, the absence of coherent splitting is evident. Numerical spectra with this background removed (Supplemental Material \cite{SupMat}, Fig.~S4) reveal that the collective decay rates are modulated by the same Bessel factor $J_n(f)$. Thus, dissipative photon exchange, which underlies superradiance and subradiance, can also be dynamically controlled, enabling on-demand tailoring of the lifetime of collective radiative modes. The behavior in the stronger-driving regime for both the HC and NHC configurations is further illustrated in Figs.~S5 and S6 of the Supplemental Material \cite{SupMat}, which show experimental data and corresponding simulations at higher pump power. In this regime, additional higher-order Floquet sidebands become visible.

To directly reveal the pump power dependence, we fix $\nu/2\pi = 11\,\mathrm{MHz}$ and sweep the pump power $P_\mathrm{pump}$ [\figpanels{fig:2}{e}{h}]. The Floquet bands display the Bessel-function revivals and collapses. In particular, when the distant qubit $Q_D$ is driven around its node frequency [\figpanel{fig:2}{g}], the driven qubit is not directly probed, yielding a clean zeroth-band splitting that maps directly onto $J_0(f)$. The successive collapses indicated by arrows coincide with the zeros of $J_0(f)$, highlighting CDT as a robust, power-controlled phenomenon. Applying the same pump power modulation to the NHC configuration [\figand{fig:2}{f}{h}] (see numerical spectra in Supplemental Material \cite{SupMat}, Fig.~S7) shows that the correlated decay can be substantially modulated, from enhanced collective emission to strong suppression.

\begin{figure}[th]
\includegraphics[width=\linewidth]{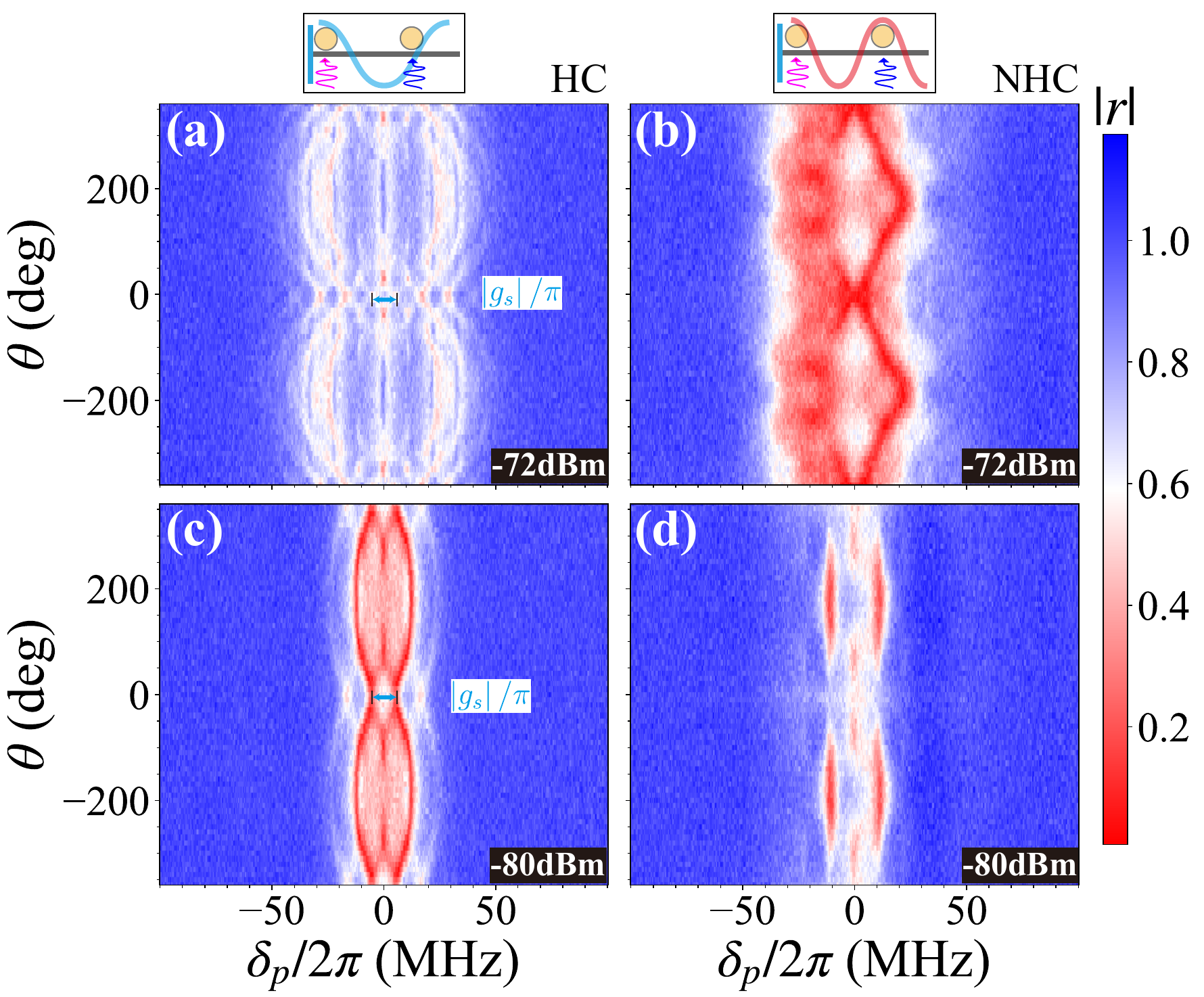}
\caption{Floquet-engineered qubit-qubit coupling as a function of the relative pump phase $\theta$. The coupling mechanism is illustrated in \figpanel{fig:1}{e}. We measure the reflection amplitude $\left | r\right |$ as a function of probe detuning $\delta_p$ and $\theta$ at a fixed modulation frequency $\nu/2\pi=\unit[11]{MHz}$. Both qubits are driven by flux pumps with a controlled relative phase $\theta$. Panels (a) and (c) are measured at the node frequency of $Q_D$, where Floquet bands split owing to virtual photon exchange, with the detailed splitting structure shown in Figs.~S8(e) and S8(g) of the Supplemental Material \cite{SupMat}.
Panels (b) and (d) are measured at the antinode frequency of $Q_D$, where qubits emit real photons into the waveguide, allowing the changes in the collective linewidths to be clearly resolved, with the detailed linewidth evolution shown in Figs.~S8(b) and S8(d) of the Supplemental Material \cite{SupMat}.
The upper panels (a), (b) correspond to a pump power $P_{pump}\approx\unit[-72]{dBm}$, and the lower panels (c), (d) correspond to $P_{pump}\approx\unit[-80]{dBm}$. 
}

\label{fig:4}
\end{figure}

A phase control knob, $\theta = \theta_d - \theta_m$, is introduced when both qubits are driven [\figpanel{fig:1}{e}]. The inter-qubit coupling then becomes (Supplemental Material \cite{SupMat}, Sec.~S4B)
\begin{equation}
g_n = g_s\,J_n\bigl[2f\sin(\theta/2)\bigr].
\label{eq:twodrives}
\end{equation}
This expression reveals that the effective interaction depends critically on the phase difference, not on the individual drive phases. This control parameter enables continuous tuning of the qubit-qubit interaction strength without altering the qubit frequencies, their spatial separation, or the pump power. When the drives are in phase ($\theta=0$), the relative modulation between the two qubits vanishes, and the static coupling is recovered. As $\theta$ varies, the argument of the Bessel function grows, leading to a controlled suppression and revival of the coupling. Figures~\figpanelNoPrefix{fig:4}{a} and \figpanelNoPrefix{fig:4}{c} show this phase-dependent HC response, where the Floquet-band splitting recovers the static value near $\theta=0$ and collapses at multiple phase values near $J_0[2f\sin(\theta/2)]=0$, in agreement with \eqref{eq:twodrives}. 
In the NHC configuration [\figand{fig:4}{b}{d}], the same phase-dependent Bessel factor modulates the collective radiative linewidth. The relative pump phase therefore provides a dynamical control knob for the nonlocal qubit-qubit interaction, enabling continuous tuning of both coherent exchange and collective decay without changing the qubit frequencies, spatial separation, or pump power.
The corresponding numerical simulations are presented in Fig.~S8 of the Supplemental Material \cite{SupMat}, which are consistent with the experimental data.


\textit{Conclusion}---We have demonstrated Floquet engineering of photon-mediated collective interactions between two remote superconducting qubits in a mirror-terminated waveguide. By exploiting the standing-wave mode structure, our single-chip device accesses both Hermitian and non-Hermitian coupling configurations simply by tuning the qubit operating frequency. A periodic flux drive dresses either configuration with Floquet sidebands, forming a Floquet ladder. The effective qubit-qubit interaction is thus continuously tunable via the frequency, amplitude, and relative phase of the drive. The coupling strength follows a Bessel-function envelope, enabling the dynamical suppression of inter-qubit coupling at the Bessel zeros and the tuning of superradiance strength in a waveguide-QED setting. The Hermitian or dissipative character of the static coupling is faithfully inherited by the driven system, offering a straightforward route for the on-demand engineering of collective quantum phenomena.

This work establishes waveguide QED with Floquet-engineered interactions as a versatile platform for both quantum technologies and fundamental physics. It enables the entanglement of remote qubits in larger qubit arrays \cite{brehm2021waveguide, bello2022spin}, and provides dynamically programmable spin-spin interactions for quantum simulation of many-body Hamiltonians \cite{zhao2022probing, nguyen2024programmable, salathe2015digital, goldman2014periodically}, including models with engineered dissipation \cite{shankar2013autonomously}. The ability to switch between Hermitian and non-Hermitian coupling simply by tuning the operating frequency, combined with Floquet control \cite{goldman2014periodically, nguyen2024programmable, zhao2022probing}, also offers itinerant-photon resources for quantum networks and remote entanglement generation \cite{kurpiers2018deterministic,shah2024stabilizing, kannan2020generating, krastanov2021optically}. More broadly, this platform opens the door to exploring driven-dissipative many-body physics, including non-equilibrium phase transitions \cite{diehl2010dynamical, fitzpatrick2017observation} and topological phenomena in open quantum systems \cite{diehl2011topology, bergholtz2021exceptional, bello2019unconventional}.

\textit{Acknowledgements.}---I.-C.H. thanks technical support from JX Quantum (www.jx-quantum.com). I.-C.H. acknowledges financial support from City University of Hong Kong through the start-up project 9610569, from the Research Grants Council of Hong Kong (Grant No. 11312322) and from Guangdong Provincial Quantum Science Strategic Initiative (GDZX2303005, GDZX2203001, and GDZX2403001). 
This work was also supported by the National Key Research and Development Program of China (Grant No. 2024YFA1408900), 
the National Natural Science Foundation of China (Grants No. 12434020, No. 12325412, No. 12604550, No. 52673402, and No. 12674085), and the Key Project of the Natural Science Program of Xinjiang Uygur Autonomous Region (Grant No. 2023D01D03).
\bibliography{main}

\begin{thebibliography}{65}%
\makeatletter
\providecommand \@ifxundefined [1]{%
 \@ifx{#1\undefined}
}%
\providecommand \@ifnum [1]{%
 \ifnum #1\expandafter \@firstoftwo
 \else \expandafter \@secondoftwo
 \fi
}%
\providecommand \@ifx [1]{%
 \ifx #1\expandafter \@firstoftwo
 \else \expandafter \@secondoftwo
 \fi
}%
\providecommand \natexlab [1]{#1}%
\providecommand \enquote  [1]{``#1''}%
\providecommand \bibnamefont  [1]{#1}%
\providecommand \bibfnamefont [1]{#1}%
\providecommand \citenamefont [1]{#1}%
\providecommand \href@noop [0]{\@secondoftwo}%
\providecommand \href [0]{\begingroup \@sanitize@url \@href}%
\providecommand \@href[1]{\@@startlink{#1}\@@href}%
\providecommand \@@href[1]{\endgroup#1\@@endlink}%
\providecommand \@sanitize@url [0]{\catcode `\\12\catcode `\$12\catcode `\&12\catcode `\#12\catcode `\^12\catcode `\_12\catcode `\%12\relax}%
\providecommand \@@startlink[1]{}%
\providecommand \@@endlink[0]{}%
\providecommand \url  [0]{\begingroup\@sanitize@url \@url }%
\providecommand \@url [1]{\endgroup\@href {#1}{\urlprefix }}%
\providecommand \urlprefix  [0]{URL }%
\providecommand \Eprint [0]{\href }%
\providecommand \doibase [0]{https://doi.org/}%
\providecommand \selectlanguage [0]{\@gobble}%
\providecommand \bibinfo  [0]{\@secondoftwo}%
\providecommand \bibfield  [0]{\@secondoftwo}%
\providecommand \translation [1]{[#1]}%
\providecommand \BibitemOpen [0]{}%
\providecommand \bibitemStop [0]{}%
\providecommand \bibitemNoStop [0]{.\EOS\space}%
\providecommand \EOS [0]{\spacefactor3000\relax}%
\providecommand \BibitemShut  [1]{\csname bibitem#1\endcsname}%
\let\auto@bib@innerbib\@empty
\bibitem [{\citenamefont {Lamb~Jr}\ and\ \citenamefont {Retherford}(1947)}]{lamb1947fine}%
  \BibitemOpen
  \bibfield  {author} {\bibinfo {author} {\bibfnamefont {W.~E.}\ \bibnamefont {Lamb~Jr}}\ and\ \bibinfo {author} {\bibfnamefont {R.~C.}\ \bibnamefont {Retherford}},\ }\bibfield  {title} {\bibinfo {title} {Fine structure of the hydrogen atom by a microwave method},\ }\href {https://doi.org/10.1103/PhysRev.72.241} {\bibfield  {journal} {\bibinfo  {journal} {Physical Review}\ }\textbf {\bibinfo {volume} {72}},\ \bibinfo {pages} {241} (\bibinfo {year} {1947})}\BibitemShut {NoStop}%
\bibitem [{\citenamefont {Van~Loo}\ \emph {et~al.}(2013)\citenamefont {Van~Loo}, \citenamefont {Fedorov}, \citenamefont {Lalumiere}, \citenamefont {Sanders}, \citenamefont {Blais},\ and\ \citenamefont {Wallraff}}]{van2013photon}%
  \BibitemOpen
  \bibfield  {author} {\bibinfo {author} {\bibfnamefont {A.~F.}\ \bibnamefont {Van~Loo}}, \bibinfo {author} {\bibfnamefont {A.}~\bibnamefont {Fedorov}}, \bibinfo {author} {\bibfnamefont {K.}~\bibnamefont {Lalumiere}}, \bibinfo {author} {\bibfnamefont {B.~C.}\ \bibnamefont {Sanders}}, \bibinfo {author} {\bibfnamefont {A.}~\bibnamefont {Blais}},\ and\ \bibinfo {author} {\bibfnamefont {A.}~\bibnamefont {Wallraff}},\ }\bibfield  {title} {\bibinfo {title} {Photon-mediated interactions between distant artificial atoms},\ }\href {https://doi.org/10.1126/science.1244324} {\bibfield  {journal} {\bibinfo  {journal} {Science}\ }\textbf {\bibinfo {volume} {342}},\ \bibinfo {pages} {1494} (\bibinfo {year} {2013})}\BibitemShut {NoStop}%
\bibitem [{\citenamefont {Wen}\ \emph {et~al.}(2019)\citenamefont {Wen}, \citenamefont {Lin}, \citenamefont {Kockum}, \citenamefont {Suri}, \citenamefont {Ian}, \citenamefont {Chen}, \citenamefont {Mao}, \citenamefont {Chiu}, \citenamefont {Delsing}, \citenamefont {Nori} \emph {et~al.}}]{wen2019large}%
  \BibitemOpen
  \bibfield  {author} {\bibinfo {author} {\bibfnamefont {P.}~\bibnamefont {Wen}}, \bibinfo {author} {\bibfnamefont {K.-T.}\ \bibnamefont {Lin}}, \bibinfo {author} {\bibfnamefont {A.}~\bibnamefont {Kockum}}, \bibinfo {author} {\bibfnamefont {B.}~\bibnamefont {Suri}}, \bibinfo {author} {\bibfnamefont {H.}~\bibnamefont {Ian}}, \bibinfo {author} {\bibfnamefont {J.}~\bibnamefont {Chen}}, \bibinfo {author} {\bibfnamefont {S.}~\bibnamefont {Mao}}, \bibinfo {author} {\bibfnamefont {C.}~\bibnamefont {Chiu}}, \bibinfo {author} {\bibfnamefont {P.}~\bibnamefont {Delsing}}, \bibinfo {author} {\bibfnamefont {F.}~\bibnamefont {Nori}}, \emph {et~al.},\ }\bibfield  {title} {\bibinfo {title} {Large collective lamb shift of two distant superconducting artificial atoms},\ }\href {https://doi.org/10.1103/PhysRevLett.123.233602} {\bibfield  {journal} {\bibinfo  {journal} {Physical review letters}\ }\textbf {\bibinfo {volume} {123}},\ \bibinfo {pages} {233602} (\bibinfo {year} {2019})}\BibitemShut {NoStop}%
\bibitem [{\citenamefont {Dicke}(1954)}]{dicke1954coherence}%
  \BibitemOpen
  \bibfield  {author} {\bibinfo {author} {\bibfnamefont {R.~H.}\ \bibnamefont {Dicke}},\ }\bibfield  {title} {\bibinfo {title} {Coherence in spontaneous radiation processes},\ }\href {https://doi.org/10.1103/PhysRev.93.99} {\bibfield  {journal} {\bibinfo  {journal} {Physical review}\ }\textbf {\bibinfo {volume} {93}},\ \bibinfo {pages} {99} (\bibinfo {year} {1954})}\BibitemShut {NoStop}%
\bibitem [{\citenamefont {Mlynek}\ \emph {et~al.}(2014)\citenamefont {Mlynek}, \citenamefont {Abdumalikov}, \citenamefont {Eichler},\ and\ \citenamefont {Wallraff}}]{mlynek2014observation}%
  \BibitemOpen
  \bibfield  {author} {\bibinfo {author} {\bibfnamefont {J.~A.}\ \bibnamefont {Mlynek}}, \bibinfo {author} {\bibfnamefont {A.~A.}\ \bibnamefont {Abdumalikov}}, \bibinfo {author} {\bibfnamefont {C.}~\bibnamefont {Eichler}},\ and\ \bibinfo {author} {\bibfnamefont {A.}~\bibnamefont {Wallraff}},\ }\bibfield  {title} {\bibinfo {title} {Observation of dicke superradiance for two artificial atoms in a cavity with high decay rate},\ }\href {https://doi.org/10.1038/ncomms6186} {\bibfield  {journal} {\bibinfo  {journal} {Nature communications}\ }\textbf {\bibinfo {volume} {5}},\ \bibinfo {pages} {5186} (\bibinfo {year} {2014})}\BibitemShut {NoStop}%
\bibitem [{\citenamefont {Zanner}\ \emph {et~al.}(2022)\citenamefont {Zanner}, \citenamefont {Orell}, \citenamefont {Schneider}, \citenamefont {Albert}, \citenamefont {Oleschko}, \citenamefont {Juan}, \citenamefont {Silveri},\ and\ \citenamefont {Kirchmair}}]{zanner2022coherent}%
  \BibitemOpen
  \bibfield  {author} {\bibinfo {author} {\bibfnamefont {M.}~\bibnamefont {Zanner}}, \bibinfo {author} {\bibfnamefont {T.}~\bibnamefont {Orell}}, \bibinfo {author} {\bibfnamefont {C.~M.}\ \bibnamefont {Schneider}}, \bibinfo {author} {\bibfnamefont {R.}~\bibnamefont {Albert}}, \bibinfo {author} {\bibfnamefont {S.}~\bibnamefont {Oleschko}}, \bibinfo {author} {\bibfnamefont {M.~L.}\ \bibnamefont {Juan}}, \bibinfo {author} {\bibfnamefont {M.}~\bibnamefont {Silveri}},\ and\ \bibinfo {author} {\bibfnamefont {G.}~\bibnamefont {Kirchmair}},\ }\bibfield  {title} {\bibinfo {title} {Coherent control of a multi-qubit dark state in waveguide quantum electrodynamics},\ }\href {https://doi.org/10.1038/s41567-022-01527-w} {\bibfield  {journal} {\bibinfo  {journal} {Nature Physics}\ }\textbf {\bibinfo {volume} {18}},\ \bibinfo {pages} {538} (\bibinfo {year} {2022})}\BibitemShut {NoStop}%
\bibitem [{\citenamefont {Bao}\ and\ \citenamefont {Lin}(2024)}]{bao2024level}%
  \BibitemOpen
  \bibfield  {author} {\bibinfo {author} {\bibfnamefont {D.}~\bibnamefont {Bao}}\ and\ \bibinfo {author} {\bibfnamefont {Z.}~\bibnamefont {Lin}},\ }\bibfield  {title} {\bibinfo {title} {Level splitting induced by collective interaction in a waveguide {QED} system},\ }\href {https://doi.org/10.1103/PhysRevResearch.6.043121} {\bibfield  {journal} {\bibinfo  {journal} {Physical Review Research}\ }\textbf {\bibinfo {volume} {6}},\ \bibinfo {pages} {043121} (\bibinfo {year} {2024})}\BibitemShut {NoStop}%
\bibitem [{\citenamefont {Sheremet}\ \emph {et~al.}(2023)\citenamefont {Sheremet}, \citenamefont {Petrov}, \citenamefont {Iorsh}, \citenamefont {Poshakinskiy},\ and\ \citenamefont {Poddubny}}]{sheremet2023waveguide}%
  \BibitemOpen
  \bibfield  {author} {\bibinfo {author} {\bibfnamefont {A.~S.}\ \bibnamefont {Sheremet}}, \bibinfo {author} {\bibfnamefont {M.~I.}\ \bibnamefont {Petrov}}, \bibinfo {author} {\bibfnamefont {I.~V.}\ \bibnamefont {Iorsh}}, \bibinfo {author} {\bibfnamefont {A.~V.}\ \bibnamefont {Poshakinskiy}},\ and\ \bibinfo {author} {\bibfnamefont {A.~N.}\ \bibnamefont {Poddubny}},\ }\bibfield  {title} {\bibinfo {title} {Waveguide quantum electrodynamics: Collective radiance and photon-photon correlations},\ }\href {https://doi.org/10.1103/RevModPhys.95.015002} {\bibfield  {journal} {\bibinfo  {journal} {Reviews of Modern Physics}\ }\textbf {\bibinfo {volume} {95}},\ \bibinfo {pages} {015002} (\bibinfo {year} {2023})}\BibitemShut {NoStop}%
\bibitem [{\citenamefont {Lalumiere}\ \emph {et~al.}(2013)\citenamefont {Lalumiere}, \citenamefont {Sanders}, \citenamefont {van Loo}, \citenamefont {Fedorov}, \citenamefont {Wallraff},\ and\ \citenamefont {Blais}}]{lalumiere2013input}%
  \BibitemOpen
  \bibfield  {author} {\bibinfo {author} {\bibfnamefont {K.}~\bibnamefont {Lalumiere}}, \bibinfo {author} {\bibfnamefont {B.~C.}\ \bibnamefont {Sanders}}, \bibinfo {author} {\bibfnamefont {A.}~\bibnamefont {van Loo}}, \bibinfo {author} {\bibfnamefont {A.}~\bibnamefont {Fedorov}}, \bibinfo {author} {\bibfnamefont {A.}~\bibnamefont {Wallraff}},\ and\ \bibinfo {author} {\bibfnamefont {A.}~\bibnamefont {Blais}},\ }\bibfield  {title} {\bibinfo {title} {Input-output theory for waveguide {QED} with an ensemble of inhomogeneous atoms},\ }\href {https://doi.org/10.1103/PhysRevA.88.043806} {\bibfield  {journal} {\bibinfo  {journal} {Physical Review A-Atomic, Molecular, and Optical Physics}\ }\textbf {\bibinfo {volume} {88}},\ \bibinfo {pages} {043806} (\bibinfo {year} {2013})}\BibitemShut {NoStop}%
\bibitem [{\citenamefont {Lin}\ \emph {et~al.}(2019)\citenamefont {Lin}, \citenamefont {Hsu}, \citenamefont {Lee}, \citenamefont {Hoi},\ and\ \citenamefont {Lin}}]{lin2019scalable}%
  \BibitemOpen
  \bibfield  {author} {\bibinfo {author} {\bibfnamefont {K.-T.}\ \bibnamefont {Lin}}, \bibinfo {author} {\bibfnamefont {T.}~\bibnamefont {Hsu}}, \bibinfo {author} {\bibfnamefont {C.-Y.}\ \bibnamefont {Lee}}, \bibinfo {author} {\bibfnamefont {I.-C.}\ \bibnamefont {Hoi}},\ and\ \bibinfo {author} {\bibfnamefont {G.-D.}\ \bibnamefont {Lin}},\ }\bibfield  {title} {\bibinfo {title} {Scalable collective lamb shift of a {1D} superconducting qubit array in front of a mirror},\ }\href {https://doi.org/10.1038/s41598-019-55545-5} {\bibfield  {journal} {\bibinfo  {journal} {Scientific reports}\ }\textbf {\bibinfo {volume} {9}},\ \bibinfo {pages} {19175} (\bibinfo {year} {2019})}\BibitemShut {NoStop}%
\bibitem [{\citenamefont {Hoi}\ \emph {et~al.}(2015)\citenamefont {Hoi}, \citenamefont {Kockum}, \citenamefont {Tornberg}, \citenamefont {Pourkabirian}, \citenamefont {Johansson}, \citenamefont {Delsing},\ and\ \citenamefont {Wilson}}]{hoi2015probing}%
  \BibitemOpen
  \bibfield  {author} {\bibinfo {author} {\bibfnamefont {I.-C.}\ \bibnamefont {Hoi}}, \bibinfo {author} {\bibfnamefont {A.~F.}\ \bibnamefont {Kockum}}, \bibinfo {author} {\bibfnamefont {L.}~\bibnamefont {Tornberg}}, \bibinfo {author} {\bibfnamefont {A.}~\bibnamefont {Pourkabirian}}, \bibinfo {author} {\bibfnamefont {G.}~\bibnamefont {Johansson}}, \bibinfo {author} {\bibfnamefont {P.}~\bibnamefont {Delsing}},\ and\ \bibinfo {author} {\bibfnamefont {C.}~\bibnamefont {Wilson}},\ }\bibfield  {title} {\bibinfo {title} {Probing the quantum vacuum with an artificial atom in front of a mirror},\ }\href {https://doi.org/10.1038/nphys3484} {\bibfield  {journal} {\bibinfo  {journal} {Nature Physics}\ }\textbf {\bibinfo {volume} {11}},\ \bibinfo {pages} {1045} (\bibinfo {year} {2015})}\BibitemShut {NoStop}%
\bibitem [{\citenamefont {Run}\ \emph {et~al.}(2026)\citenamefont {Run}, \citenamefont {Lin}, \citenamefont {Hsieh}, \citenamefont {Wu}, \citenamefont {Zhou}, \citenamefont {Lin}, \citenamefont {Kockum},\ and\ \citenamefont {Hoi}}]{run2026realizing}%
  \BibitemOpen
  \bibfield  {author} {\bibinfo {author} {\bibfnamefont {C.-X.}\ \bibnamefont {Run}}, \bibinfo {author} {\bibfnamefont {K.-T.}\ \bibnamefont {Lin}}, \bibinfo {author} {\bibfnamefont {K.-M.}\ \bibnamefont {Hsieh}}, \bibinfo {author} {\bibfnamefont {B.-Y.}\ \bibnamefont {Wu}}, \bibinfo {author} {\bibfnamefont {W.-M.}\ \bibnamefont {Zhou}}, \bibinfo {author} {\bibfnamefont {G.-D.}\ \bibnamefont {Lin}}, \bibinfo {author} {\bibfnamefont {A.~F.}\ \bibnamefont {Kockum}},\ and\ \bibinfo {author} {\bibfnamefont {I.-C.}\ \bibnamefont {Hoi}},\ }\bibfield  {title} {\bibinfo {title} {Realizing on-demand all-to-all selective interactions between distant spin ensembles},\ }\href {https://doi.org/10.1103/r4y6-8npw} {\bibfield  {journal} {\bibinfo  {journal} {Physical Review Letters}\ }\textbf {\bibinfo {volume} {136}},\ \bibinfo {pages} {206905} (\bibinfo {year} {2026})}\BibitemShut {NoStop}%
\bibitem [{Sup()}]{SupMat}%
  \BibitemOpen
  \href@noop {} {}\bibinfo {note} {See Supplemental Material at [URL] for details of the experimental setup, device and qubit characterization, static Hermitian and non-Hermitian qubit-qubit interactions, the master-equation calculation of the reflection response, and the derivation of the effective Floquet Hamiltonians. Additional measurements and simulations are presented for the dependence on modulation frequency, pump power, and relative pump phase. The discussions in the material include Refs.~ \cite{koch2007charge, probst2015efficient, lin2019scalable, lu2021characterizing, wen2019large, van2013photon, bao2024level, cheng2024tuning}.}\BibitemShut {Stop}%
\bibitem [{\citenamefont {Goldman}\ and\ \citenamefont {Dalibard}(2014)}]{goldman2014periodically}%
  \BibitemOpen
  \bibfield  {author} {\bibinfo {author} {\bibfnamefont {N.}~\bibnamefont {Goldman}}\ and\ \bibinfo {author} {\bibfnamefont {J.}~\bibnamefont {Dalibard}},\ }\bibfield  {title} {\bibinfo {title} {Periodically driven quantum systems: Effective {Hamiltonians} and engineered gauge fields},\ }\href {https://doi.org/10.1103/PhysRevX.4.031027} {\bibfield  {journal} {\bibinfo  {journal} {Physical review X}\ }\textbf {\bibinfo {volume} {4}},\ \bibinfo {pages} {031027} (\bibinfo {year} {2014})}\BibitemShut {NoStop}%
\bibitem [{\citenamefont {Jotzu}\ \emph {et~al.}(2014)\citenamefont {Jotzu}, \citenamefont {Messer}, \citenamefont {Desbuquois}, \citenamefont {Lebrat}, \citenamefont {Uehlinger}, \citenamefont {Greif},\ and\ \citenamefont {Esslinger}}]{jotzu2014experimental}%
  \BibitemOpen
  \bibfield  {author} {\bibinfo {author} {\bibfnamefont {G.}~\bibnamefont {Jotzu}}, \bibinfo {author} {\bibfnamefont {M.}~\bibnamefont {Messer}}, \bibinfo {author} {\bibfnamefont {R.}~\bibnamefont {Desbuquois}}, \bibinfo {author} {\bibfnamefont {M.}~\bibnamefont {Lebrat}}, \bibinfo {author} {\bibfnamefont {T.}~\bibnamefont {Uehlinger}}, \bibinfo {author} {\bibfnamefont {D.}~\bibnamefont {Greif}},\ and\ \bibinfo {author} {\bibfnamefont {T.}~\bibnamefont {Esslinger}},\ }\bibfield  {title} {\bibinfo {title} {Experimental realization of the topological haldane model with ultracold fermions},\ }\href {https://doi.org/10.1038/nature13915} {\bibfield  {journal} {\bibinfo  {journal} {Nature}\ }\textbf {\bibinfo {volume} {515}},\ \bibinfo {pages} {237} (\bibinfo {year} {2014})}\BibitemShut {NoStop}%
\bibitem [{\citenamefont {Wang}\ \emph {et~al.}(2024)\citenamefont {Wang}, \citenamefont {Zhang}, \citenamefont {Tang},\ and\ \citenamefont {Wang}}]{wang2024floquet}%
  \BibitemOpen
  \bibfield  {author} {\bibinfo {author} {\bibfnamefont {W.}~\bibnamefont {Wang}}, \bibinfo {author} {\bibfnamefont {Z.}~\bibnamefont {Zhang}}, \bibinfo {author} {\bibfnamefont {G.-X.}\ \bibnamefont {Tang}},\ and\ \bibinfo {author} {\bibfnamefont {T.}~\bibnamefont {Wang}},\ }\bibfield  {title} {\bibinfo {title} {Floquet engineering tunable periodic gauge fields and simulating real topological phases in a cold-alkaline-earth-metal-atom optical lattice},\ }\href {https://doi.org/10.1103/PhysRevA.110.023308} {\bibfield  {journal} {\bibinfo  {journal} {Physical Review A}\ }\textbf {\bibinfo {volume} {110}},\ \bibinfo {pages} {023308} (\bibinfo {year} {2024})}\BibitemShut {NoStop}%
\bibitem [{\citenamefont {An}\ \emph {et~al.}(2017)\citenamefont {An}, \citenamefont {Meier},\ and\ \citenamefont {Gadway}}]{an2017direct}%
  \BibitemOpen
  \bibfield  {author} {\bibinfo {author} {\bibfnamefont {F.~A.}\ \bibnamefont {An}}, \bibinfo {author} {\bibfnamefont {E.~J.}\ \bibnamefont {Meier}},\ and\ \bibinfo {author} {\bibfnamefont {B.}~\bibnamefont {Gadway}},\ }\bibfield  {title} {\bibinfo {title} {Direct observation of chiral currents and magnetic reflection in atomic flux lattices},\ }\href {https://doi.org/10.1126/sciadv.1602685} {\bibfield  {journal} {\bibinfo  {journal} {Science advances}\ }\textbf {\bibinfo {volume} {3}},\ \bibinfo {pages} {e1602685} (\bibinfo {year} {2017})}\BibitemShut {NoStop}%
\bibitem [{\citenamefont {Xu}\ \emph {et~al.}(2022)\citenamefont {Xu}, \citenamefont {Wang}, \citenamefont {Dai}, \citenamefont {Mao}, \citenamefont {Cai}, \citenamefont {Zhu},\ and\ \citenamefont {Wang}}]{xu2022floquet}%
  \BibitemOpen
  \bibfield  {author} {\bibinfo {author} {\bibfnamefont {X.}~\bibnamefont {Xu}}, \bibinfo {author} {\bibfnamefont {J.}~\bibnamefont {Wang}}, \bibinfo {author} {\bibfnamefont {J.}~\bibnamefont {Dai}}, \bibinfo {author} {\bibfnamefont {R.}~\bibnamefont {Mao}}, \bibinfo {author} {\bibfnamefont {H.}~\bibnamefont {Cai}}, \bibinfo {author} {\bibfnamefont {S.-Y.}\ \bibnamefont {Zhu}},\ and\ \bibinfo {author} {\bibfnamefont {D.-W.}\ \bibnamefont {Wang}},\ }\bibfield  {title} {\bibinfo {title} {Floquet superradiance lattices in thermal atoms},\ }\href {https://doi.org/10.1103/PhysRevLett.129.273603} {\bibfield  {journal} {\bibinfo  {journal} {Phys. Rev. Lett.}\ }\textbf {\bibinfo {volume} {129}},\ \bibinfo {pages} {273603} (\bibinfo {year} {2022})}\BibitemShut {NoStop}%
\bibitem [{\citenamefont {Wang}\ \emph {et~al.}(2015)\citenamefont {Wang}, \citenamefont {Cai}, \citenamefont {Yuan}, \citenamefont {Zhu},\ and\ \citenamefont {Liu}}]{wang2015topological}%
  \BibitemOpen
  \bibfield  {author} {\bibinfo {author} {\bibfnamefont {D.-W.}\ \bibnamefont {Wang}}, \bibinfo {author} {\bibfnamefont {H.}~\bibnamefont {Cai}}, \bibinfo {author} {\bibfnamefont {L.}~\bibnamefont {Yuan}}, \bibinfo {author} {\bibfnamefont {S.-Y.}\ \bibnamefont {Zhu}},\ and\ \bibinfo {author} {\bibfnamefont {R.-B.}\ \bibnamefont {Liu}},\ }\bibfield  {title} {\bibinfo {title} {Topological phase transitions in superradiance lattices},\ }\href {https://opg.optica.org/optica/abstract.cfm?URI=optica-2-8-712} {\bibfield  {journal} {\bibinfo  {journal} {Optica}\ }\textbf {\bibinfo {volume} {2}},\ \bibinfo {pages} {712} (\bibinfo {year} {2015})}\BibitemShut {NoStop}%
\bibitem [{\citenamefont {Rechtsman}\ \emph {et~al.}(2013)\citenamefont {Rechtsman}, \citenamefont {Zeuner}, \citenamefont {Plotnik}, \citenamefont {Lumer}, \citenamefont {Podolsky}, \citenamefont {Dreisow}, \citenamefont {Nolte}, \citenamefont {Segev},\ and\ \citenamefont {Szameit}}]{rechtsman2013photonic}%
  \BibitemOpen
  \bibfield  {author} {\bibinfo {author} {\bibfnamefont {M.~C.}\ \bibnamefont {Rechtsman}}, \bibinfo {author} {\bibfnamefont {J.~M.}\ \bibnamefont {Zeuner}}, \bibinfo {author} {\bibfnamefont {Y.}~\bibnamefont {Plotnik}}, \bibinfo {author} {\bibfnamefont {Y.}~\bibnamefont {Lumer}}, \bibinfo {author} {\bibfnamefont {D.}~\bibnamefont {Podolsky}}, \bibinfo {author} {\bibfnamefont {F.}~\bibnamefont {Dreisow}}, \bibinfo {author} {\bibfnamefont {S.}~\bibnamefont {Nolte}}, \bibinfo {author} {\bibfnamefont {M.}~\bibnamefont {Segev}},\ and\ \bibinfo {author} {\bibfnamefont {A.}~\bibnamefont {Szameit}},\ }\bibfield  {title} {\bibinfo {title} {Photonic floquet topological insulators},\ }\href {https://doi.org/10.1038/nature12066} {\bibfield  {journal} {\bibinfo  {journal} {Nature}\ }\textbf {\bibinfo {volume} {496}},\ \bibinfo {pages} {196} (\bibinfo {year} {2013})}\BibitemShut {NoStop}%
\bibitem [{\citenamefont {McIver}\ \emph {et~al.}(2020)\citenamefont {McIver}, \citenamefont {Schulte}, \citenamefont {Stein}, \citenamefont {Matsuyama}, \citenamefont {Jotzu}, \citenamefont {Meier},\ and\ \citenamefont {Cavalleri}}]{mciver2020light}%
  \BibitemOpen
  \bibfield  {author} {\bibinfo {author} {\bibfnamefont {J.~W.}\ \bibnamefont {McIver}}, \bibinfo {author} {\bibfnamefont {B.}~\bibnamefont {Schulte}}, \bibinfo {author} {\bibfnamefont {F.-U.}\ \bibnamefont {Stein}}, \bibinfo {author} {\bibfnamefont {T.}~\bibnamefont {Matsuyama}}, \bibinfo {author} {\bibfnamefont {G.}~\bibnamefont {Jotzu}}, \bibinfo {author} {\bibfnamefont {G.}~\bibnamefont {Meier}},\ and\ \bibinfo {author} {\bibfnamefont {A.}~\bibnamefont {Cavalleri}},\ }\bibfield  {title} {\bibinfo {title} {Light-induced anomalous hall effect in graphene},\ }\href {https://doi.org/10.1038/s41567-019-0698-y} {\bibfield  {journal} {\bibinfo  {journal} {Nature physics}\ }\textbf {\bibinfo {volume} {16}},\ \bibinfo {pages} {38} (\bibinfo {year} {2020})}\BibitemShut {NoStop}%
\bibitem [{\citenamefont {Maczewsky}\ \emph {et~al.}(2017)\citenamefont {Maczewsky}, \citenamefont {Zeuner}, \citenamefont {Nolte},\ and\ \citenamefont {Szameit}}]{maczewsky2017observation}%
  \BibitemOpen
  \bibfield  {author} {\bibinfo {author} {\bibfnamefont {L.~J.}\ \bibnamefont {Maczewsky}}, \bibinfo {author} {\bibfnamefont {J.~M.}\ \bibnamefont {Zeuner}}, \bibinfo {author} {\bibfnamefont {S.}~\bibnamefont {Nolte}},\ and\ \bibinfo {author} {\bibfnamefont {A.}~\bibnamefont {Szameit}},\ }\bibfield  {title} {\bibinfo {title} {Observation of photonic anomalous floquet topological insulators},\ }\href {https://doi.org/10.1038/ncomms13756} {\bibfield  {journal} {\bibinfo  {journal} {Nature communications}\ }\textbf {\bibinfo {volume} {8}},\ \bibinfo {pages} {13756} (\bibinfo {year} {2017})}\BibitemShut {NoStop}%
\bibitem [{\citenamefont {Afzal}\ \emph {et~al.}(2020)\citenamefont {Afzal}, \citenamefont {Zimmerling}, \citenamefont {Ren}, \citenamefont {Perron},\ and\ \citenamefont {Van}}]{afzal2020realization}%
  \BibitemOpen
  \bibfield  {author} {\bibinfo {author} {\bibfnamefont {S.}~\bibnamefont {Afzal}}, \bibinfo {author} {\bibfnamefont {T.~J.}\ \bibnamefont {Zimmerling}}, \bibinfo {author} {\bibfnamefont {Y.}~\bibnamefont {Ren}}, \bibinfo {author} {\bibfnamefont {D.}~\bibnamefont {Perron}},\ and\ \bibinfo {author} {\bibfnamefont {V.}~\bibnamefont {Van}},\ }\bibfield  {title} {\bibinfo {title} {Realization of anomalous floquet insulators in strongly coupled nanophotonic lattices},\ }\href {https://doi.org/10.1103/PhysRevLett.124.253601} {\bibfield  {journal} {\bibinfo  {journal} {Physical Review Letters}\ }\textbf {\bibinfo {volume} {124}},\ \bibinfo {pages} {253601} (\bibinfo {year} {2020})}\BibitemShut {NoStop}%
\bibitem [{\citenamefont {Oka}\ and\ \citenamefont {Aoki}(2009)}]{oka2009photovoltaic}%
  \BibitemOpen
  \bibfield  {author} {\bibinfo {author} {\bibfnamefont {T.}~\bibnamefont {Oka}}\ and\ \bibinfo {author} {\bibfnamefont {H.}~\bibnamefont {Aoki}},\ }\bibfield  {title} {\bibinfo {title} {Photovoltaic hall effect in graphene},\ }\href {https://doi.org/10.1103/PhysRevB.79.081406} {\bibfield  {journal} {\bibinfo  {journal} {Physical Review B-Condensed Matter and Materials Physics}\ }\textbf {\bibinfo {volume} {79}},\ \bibinfo {pages} {081406} (\bibinfo {year} {2009})}\BibitemShut {NoStop}%
\bibitem [{\citenamefont {Deng}\ \emph {et~al.}(2022)\citenamefont {Deng}, \citenamefont {Dong}, \citenamefont {Zhang}, \citenamefont {Wu}, \citenamefont {Yuan}, \citenamefont {Zhu}, \citenamefont {Jin}, \citenamefont {Li}, \citenamefont {Wang}, \citenamefont {Cai}, \citenamefont {Song}, \citenamefont {Wang}, \citenamefont {You},\ and\ \citenamefont {Wang}}]{deng2022}%
  \BibitemOpen
  \bibfield  {author} {\bibinfo {author} {\bibfnamefont {J.}~\bibnamefont {Deng}}, \bibinfo {author} {\bibfnamefont {H.}~\bibnamefont {Dong}}, \bibinfo {author} {\bibfnamefont {C.}~\bibnamefont {Zhang}}, \bibinfo {author} {\bibfnamefont {Y.}~\bibnamefont {Wu}}, \bibinfo {author} {\bibfnamefont {J.}~\bibnamefont {Yuan}}, \bibinfo {author} {\bibfnamefont {X.}~\bibnamefont {Zhu}}, \bibinfo {author} {\bibfnamefont {F.}~\bibnamefont {Jin}}, \bibinfo {author} {\bibfnamefont {H.}~\bibnamefont {Li}}, \bibinfo {author} {\bibfnamefont {Z.}~\bibnamefont {Wang}}, \bibinfo {author} {\bibfnamefont {H.}~\bibnamefont {Cai}}, \bibinfo {author} {\bibfnamefont {C.}~\bibnamefont {Song}}, \bibinfo {author} {\bibfnamefont {H.}~\bibnamefont {Wang}}, \bibinfo {author} {\bibfnamefont {J.~Q.}\ \bibnamefont {You}},\ and\ \bibinfo {author} {\bibfnamefont {D.-W.}\ \bibnamefont {Wang}},\ }\bibfield  {title} {\bibinfo {title} {Observing the quantum topology of light},\ }\href {https://doi.org/10.1126/science.ade6219} {\bibfield  {journal}
  {\bibinfo  {journal} {Science}\ }\textbf {\bibinfo {volume} {378}},\ \bibinfo {pages} {966} (\bibinfo {year} {2022})}\BibitemShut {NoStop}%
\bibitem [{\citenamefont {Xu}\ \emph {et~al.}(2020)\citenamefont {Xu}, \citenamefont {Zhong}, \citenamefont {Han}, \citenamefont {Jin}, \citenamefont {Jiang},\ and\ \citenamefont {Zhang}}]{xu2020floquet}%
  \BibitemOpen
  \bibfield  {author} {\bibinfo {author} {\bibfnamefont {J.}~\bibnamefont {Xu}}, \bibinfo {author} {\bibfnamefont {C.}~\bibnamefont {Zhong}}, \bibinfo {author} {\bibfnamefont {X.}~\bibnamefont {Han}}, \bibinfo {author} {\bibfnamefont {D.}~\bibnamefont {Jin}}, \bibinfo {author} {\bibfnamefont {L.}~\bibnamefont {Jiang}},\ and\ \bibinfo {author} {\bibfnamefont {X.}~\bibnamefont {Zhang}},\ }\bibfield  {title} {\bibinfo {title} {Floquet cavity electromagnonics},\ }\href {https://doi.org/10.1103/PhysRevLett.125.237201} {\bibfield  {journal} {\bibinfo  {journal} {Physical Review Letters}\ }\textbf {\bibinfo {volume} {125}},\ \bibinfo {pages} {237201} (\bibinfo {year} {2020})}\BibitemShut {NoStop}%
\bibitem [{\citenamefont {Zhang}\ \emph {et~al.}(2017)\citenamefont {Zhang}, \citenamefont {Hess}, \citenamefont {Kyprianidis}, \citenamefont {Becker}, \citenamefont {Lee}, \citenamefont {Smith}, \citenamefont {Pagano}, \citenamefont {Potirniche}, \citenamefont {Potter}, \citenamefont {Vishwanath} \emph {et~al.}}]{zhang2017observation}%
  \BibitemOpen
  \bibfield  {author} {\bibinfo {author} {\bibfnamefont {J.}~\bibnamefont {Zhang}}, \bibinfo {author} {\bibfnamefont {P.~W.}\ \bibnamefont {Hess}}, \bibinfo {author} {\bibfnamefont {A.}~\bibnamefont {Kyprianidis}}, \bibinfo {author} {\bibfnamefont {P.}~\bibnamefont {Becker}}, \bibinfo {author} {\bibfnamefont {A.}~\bibnamefont {Lee}}, \bibinfo {author} {\bibfnamefont {J.}~\bibnamefont {Smith}}, \bibinfo {author} {\bibfnamefont {G.}~\bibnamefont {Pagano}}, \bibinfo {author} {\bibfnamefont {I.-D.}\ \bibnamefont {Potirniche}}, \bibinfo {author} {\bibfnamefont {A.~C.}\ \bibnamefont {Potter}}, \bibinfo {author} {\bibfnamefont {A.}~\bibnamefont {Vishwanath}}, \emph {et~al.},\ }\bibfield  {title} {\bibinfo {title} {Observation of a discrete time crystal},\ }\href {https://doi.org/10.1038/nature21413} {\bibfield  {journal} {\bibinfo  {journal} {Nature}\ }\textbf {\bibinfo {volume} {543}},\ \bibinfo {pages} {217} (\bibinfo {year} {2017})}\BibitemShut {NoStop}%
\bibitem [{\citenamefont {Zhang}\ \emph {et~al.}(2022{\natexlab{a}})\citenamefont {Zhang}, \citenamefont {Jiang}, \citenamefont {Deng}, \citenamefont {Wang}, \citenamefont {Chen}, \citenamefont {Zhang}, \citenamefont {Ren}, \citenamefont {Dong}, \citenamefont {Xu}, \citenamefont {Gao} \emph {et~al.}}]{zhang2022digital}%
  \BibitemOpen
  \bibfield  {author} {\bibinfo {author} {\bibfnamefont {X.}~\bibnamefont {Zhang}}, \bibinfo {author} {\bibfnamefont {W.}~\bibnamefont {Jiang}}, \bibinfo {author} {\bibfnamefont {J.}~\bibnamefont {Deng}}, \bibinfo {author} {\bibfnamefont {K.}~\bibnamefont {Wang}}, \bibinfo {author} {\bibfnamefont {J.}~\bibnamefont {Chen}}, \bibinfo {author} {\bibfnamefont {P.}~\bibnamefont {Zhang}}, \bibinfo {author} {\bibfnamefont {W.}~\bibnamefont {Ren}}, \bibinfo {author} {\bibfnamefont {H.}~\bibnamefont {Dong}}, \bibinfo {author} {\bibfnamefont {S.}~\bibnamefont {Xu}}, \bibinfo {author} {\bibfnamefont {Y.}~\bibnamefont {Gao}}, \emph {et~al.},\ }\bibfield  {title} {\bibinfo {title} {{Digital quantum simulation of Floquet symmetry-protected topological phases}},\ }\href {https://doi.org/https://doi.org/10.1038/s41586-022-04854-3} {\bibfield  {journal} {\bibinfo  {journal} {Nature}\ }\textbf {\bibinfo {volume} {607}},\ \bibinfo {pages} {468} (\bibinfo {year} {2022}{\natexlab{a}})}\BibitemShut {NoStop}%
\bibitem [{\citenamefont {Dumitrescu}\ \emph {et~al.}(2022)\citenamefont {Dumitrescu}, \citenamefont {Bohnet}, \citenamefont {Gaebler}, \citenamefont {Hankin}, \citenamefont {Hayes}, \citenamefont {Kumar}, \citenamefont {Neyenhuis}, \citenamefont {Vasseur},\ and\ \citenamefont {Potter}}]{dumitrescu2022dynamical}%
  \BibitemOpen
  \bibfield  {author} {\bibinfo {author} {\bibfnamefont {P.~T.}\ \bibnamefont {Dumitrescu}}, \bibinfo {author} {\bibfnamefont {J.~G.}\ \bibnamefont {Bohnet}}, \bibinfo {author} {\bibfnamefont {J.~P.}\ \bibnamefont {Gaebler}}, \bibinfo {author} {\bibfnamefont {A.}~\bibnamefont {Hankin}}, \bibinfo {author} {\bibfnamefont {D.}~\bibnamefont {Hayes}}, \bibinfo {author} {\bibfnamefont {A.}~\bibnamefont {Kumar}}, \bibinfo {author} {\bibfnamefont {B.}~\bibnamefont {Neyenhuis}}, \bibinfo {author} {\bibfnamefont {R.}~\bibnamefont {Vasseur}},\ and\ \bibinfo {author} {\bibfnamefont {A.~C.}\ \bibnamefont {Potter}},\ }\bibfield  {title} {\bibinfo {title} {Dynamical topological phase realized in a trapped-ion quantum simulator},\ }\href {https://doi.org/10.1038/s41586-022-04853-4} {\bibfield  {journal} {\bibinfo  {journal} {Nature}\ }\textbf {\bibinfo {volume} {607}},\ \bibinfo {pages} {463} (\bibinfo {year} {2022})}\BibitemShut {NoStop}%
\bibitem [{\citenamefont {Kiefer}\ \emph {et~al.}(2019)\citenamefont {Kiefer}, \citenamefont {Hakelberg}, \citenamefont {Wittemer}, \citenamefont {Berm{\'u}dez}, \citenamefont {Porras}, \citenamefont {Warring},\ and\ \citenamefont {Schaetz}}]{kiefer2019floquet}%
  \BibitemOpen
  \bibfield  {author} {\bibinfo {author} {\bibfnamefont {P.}~\bibnamefont {Kiefer}}, \bibinfo {author} {\bibfnamefont {F.}~\bibnamefont {Hakelberg}}, \bibinfo {author} {\bibfnamefont {M.}~\bibnamefont {Wittemer}}, \bibinfo {author} {\bibfnamefont {A.}~\bibnamefont {Berm{\'u}dez}}, \bibinfo {author} {\bibfnamefont {D.}~\bibnamefont {Porras}}, \bibinfo {author} {\bibfnamefont {U.}~\bibnamefont {Warring}},\ and\ \bibinfo {author} {\bibfnamefont {T.}~\bibnamefont {Schaetz}},\ }\bibfield  {title} {\bibinfo {title} {Floquet-engineered vibrational dynamics in a two-dimensional array of trapped ions},\ }\href {https://doi.org/10.1103/PhysRevLett.123.213605} {\bibfield  {journal} {\bibinfo  {journal} {Physical Review Letters}\ }\textbf {\bibinfo {volume} {123}},\ \bibinfo {pages} {213605} (\bibinfo {year} {2019})}\BibitemShut {NoStop}%
\bibitem [{\citenamefont {Nguyen}\ \emph {et~al.}(2024)\citenamefont {Nguyen}, \citenamefont {Kim}, \citenamefont {Hashim}, \citenamefont {Goss}, \citenamefont {Marinelli}, \citenamefont {Bhandari}, \citenamefont {Das}, \citenamefont {Naik}, \citenamefont {Kreikebaum}, \citenamefont {Jordan} \emph {et~al.}}]{nguyen2024programmable}%
  \BibitemOpen
  \bibfield  {author} {\bibinfo {author} {\bibfnamefont {L.~B.}\ \bibnamefont {Nguyen}}, \bibinfo {author} {\bibfnamefont {Y.}~\bibnamefont {Kim}}, \bibinfo {author} {\bibfnamefont {A.}~\bibnamefont {Hashim}}, \bibinfo {author} {\bibfnamefont {N.}~\bibnamefont {Goss}}, \bibinfo {author} {\bibfnamefont {B.}~\bibnamefont {Marinelli}}, \bibinfo {author} {\bibfnamefont {B.}~\bibnamefont {Bhandari}}, \bibinfo {author} {\bibfnamefont {D.}~\bibnamefont {Das}}, \bibinfo {author} {\bibfnamefont {R.~K.}\ \bibnamefont {Naik}}, \bibinfo {author} {\bibfnamefont {J.~M.}\ \bibnamefont {Kreikebaum}}, \bibinfo {author} {\bibfnamefont {A.~N.}\ \bibnamefont {Jordan}}, \emph {et~al.},\ }\bibfield  {title} {\bibinfo {title} {Programmable heisenberg interactions between floquet qubits},\ }\href {https://doi.org/10.1038/s41567-023-02326-7} {\bibfield  {journal} {\bibinfo  {journal} {Nature Physics}\ }\textbf {\bibinfo {volume} {20}},\ \bibinfo {pages} {240} (\bibinfo {year} {2024})}\BibitemShut {NoStop}%
\bibitem [{\citenamefont {Liu}\ \emph {et~al.}(2020)\citenamefont {Liu}, \citenamefont {Feng}, \citenamefont {Ren}, \citenamefont {Wang},\ and\ \citenamefont {Wang}}]{liu2020synthesizing}%
  \BibitemOpen
  \bibfield  {author} {\bibinfo {author} {\bibfnamefont {W.}~\bibnamefont {Liu}}, \bibinfo {author} {\bibfnamefont {W.}~\bibnamefont {Feng}}, \bibinfo {author} {\bibfnamefont {W.}~\bibnamefont {Ren}}, \bibinfo {author} {\bibfnamefont {D.-W.}\ \bibnamefont {Wang}},\ and\ \bibinfo {author} {\bibfnamefont {H.}~\bibnamefont {Wang}},\ }\bibfield  {title} {\bibinfo {title} {Synthesizing three-body interaction of spin chirality with superconducting qubits},\ }\href {https://doi.org/10.1063/1.5140884} {\bibfield  {journal} {\bibinfo  {journal} {Applied physics letters}\ }\textbf {\bibinfo {volume} {116}} (\bibinfo {year} {2020})}\BibitemShut {NoStop}%
\bibitem [{\citenamefont {Rudner}\ and\ \citenamefont {Lindner}(2020)}]{rudner2020band}%
  \BibitemOpen
  \bibfield  {author} {\bibinfo {author} {\bibfnamefont {M.~S.}\ \bibnamefont {Rudner}}\ and\ \bibinfo {author} {\bibfnamefont {N.~H.}\ \bibnamefont {Lindner}},\ }\bibfield  {title} {\bibinfo {title} {Band structure engineering and non-equilibrium dynamics in floquet topological insulators},\ }\href {https://doi.org/10.1038/s42254-020-0170-z} {\bibfield  {journal} {\bibinfo  {journal} {Nature reviews physics}\ }\textbf {\bibinfo {volume} {2}},\ \bibinfo {pages} {229} (\bibinfo {year} {2020})}\BibitemShut {NoStop}%
\bibitem [{\citenamefont {Wang}\ \emph {et~al.}(2025)\citenamefont {Wang}, \citenamefont {Dai}, \citenamefont {Mao}, \citenamefont {Lu}, \citenamefont {Liu}, \citenamefont {Hu}, \citenamefont {Zhu}, \citenamefont {Xu}, \citenamefont {Cai},\ and\ \citenamefont {Wang}}]{wang2025realizing}%
  \BibitemOpen
  \bibfield  {author} {\bibinfo {author} {\bibfnamefont {J.}~\bibnamefont {Wang}}, \bibinfo {author} {\bibfnamefont {J.}~\bibnamefont {Dai}}, \bibinfo {author} {\bibfnamefont {R.}~\bibnamefont {Mao}}, \bibinfo {author} {\bibfnamefont {Y.}~\bibnamefont {Lu}}, \bibinfo {author} {\bibfnamefont {X.}~\bibnamefont {Liu}}, \bibinfo {author} {\bibfnamefont {H.}~\bibnamefont {Hu}}, \bibinfo {author} {\bibfnamefont {S.-Y.}\ \bibnamefont {Zhu}}, \bibinfo {author} {\bibfnamefont {X.}~\bibnamefont {Xu}}, \bibinfo {author} {\bibfnamefont {H.}~\bibnamefont {Cai}},\ and\ \bibinfo {author} {\bibfnamefont {D.-W.}\ \bibnamefont {Wang}},\ }\bibfield  {title} {\bibinfo {title} {Realizing the haldane model in thermal atoms},\ }\href {https://doi.org/10.48550/arXiv.2509.08411} {\bibfield  {journal} {\bibinfo  {journal} {arXiv preprint arXiv:2509.08411}\ } (\bibinfo {year} {2025})}\BibitemShut {NoStop}%
\bibitem [{\citenamefont {Petiziol}\ and\ \citenamefont {Eckardt}(2022)}]{petiziol2022cavity}%
  \BibitemOpen
  \bibfield  {author} {\bibinfo {author} {\bibfnamefont {F.}~\bibnamefont {Petiziol}}\ and\ \bibinfo {author} {\bibfnamefont {A.}~\bibnamefont {Eckardt}},\ }\bibfield  {title} {\bibinfo {title} {Cavity-based reservoir engineering for floquet-engineered superconducting circuits},\ }\href {https://doi.org/10.1103/PhysRevLett.129.233601} {\bibfield  {journal} {\bibinfo  {journal} {Physical Review Letters}\ }\textbf {\bibinfo {volume} {129}},\ \bibinfo {pages} {233601} (\bibinfo {year} {2022})}\BibitemShut {NoStop}%
\bibitem [{\citenamefont {Harrington}\ \emph {et~al.}(2022)\citenamefont {Harrington}, \citenamefont {Mueller},\ and\ \citenamefont {Murch}}]{harrington2022engineered}%
  \BibitemOpen
  \bibfield  {author} {\bibinfo {author} {\bibfnamefont {P.~M.}\ \bibnamefont {Harrington}}, \bibinfo {author} {\bibfnamefont {E.~J.}\ \bibnamefont {Mueller}},\ and\ \bibinfo {author} {\bibfnamefont {K.~W.}\ \bibnamefont {Murch}},\ }\bibfield  {title} {\bibinfo {title} {Engineered dissipation for quantum information science},\ }\href {https://doi.org/10.1038/s42254-022-00494-8} {\bibfield  {journal} {\bibinfo  {journal} {Nature Reviews Physics}\ }\textbf {\bibinfo {volume} {4}},\ \bibinfo {pages} {660} (\bibinfo {year} {2022})}\BibitemShut {NoStop}%
\bibitem [{\citenamefont {Lin}\ \emph {et~al.}(2022)\citenamefont {Lin}, \citenamefont {Lu}, \citenamefont {Wen}, \citenamefont {Cheng}, \citenamefont {Lee}, \citenamefont {Lin}, \citenamefont {Chiang}, \citenamefont {Hsieh}, \citenamefont {Chen}, \citenamefont {Chien} \emph {et~al.}}]{lin2022deterministic}%
  \BibitemOpen
  \bibfield  {author} {\bibinfo {author} {\bibfnamefont {W.-J.}\ \bibnamefont {Lin}}, \bibinfo {author} {\bibfnamefont {Y.}~\bibnamefont {Lu}}, \bibinfo {author} {\bibfnamefont {P.~Y.}\ \bibnamefont {Wen}}, \bibinfo {author} {\bibfnamefont {Y.-T.}\ \bibnamefont {Cheng}}, \bibinfo {author} {\bibfnamefont {C.-P.}\ \bibnamefont {Lee}}, \bibinfo {author} {\bibfnamefont {K.~T.}\ \bibnamefont {Lin}}, \bibinfo {author} {\bibfnamefont {K.~H.}\ \bibnamefont {Chiang}}, \bibinfo {author} {\bibfnamefont {M.~C.}\ \bibnamefont {Hsieh}}, \bibinfo {author} {\bibfnamefont {C.-Y.}\ \bibnamefont {Chen}}, \bibinfo {author} {\bibfnamefont {C.-H.}\ \bibnamefont {Chien}}, \emph {et~al.},\ }\bibfield  {title} {\bibinfo {title} {Deterministic loading of microwaves onto an artificial atom using a time-reversed waveform},\ }\href {https://doi.org/10.1021/acs.nanolett.2c02578} {\bibfield  {journal} {\bibinfo  {journal} {Nano Letters}\ }\textbf {\bibinfo {volume} {22}},\ \bibinfo {pages} {8137} (\bibinfo {year} {2022})}\BibitemShut
  {NoStop}%
\bibitem [{\citenamefont {Cheng}\ \emph {et~al.}(2024)\citenamefont {Cheng}, \citenamefont {Chien}, \citenamefont {Hsieh}, \citenamefont {Huang}, \citenamefont {Wen}, \citenamefont {Lin}, \citenamefont {Lu}, \citenamefont {Aziz}, \citenamefont {Lee}, \citenamefont {Lin} \emph {et~al.}}]{cheng2024tuning}%
  \BibitemOpen
  \bibfield  {author} {\bibinfo {author} {\bibfnamefont {Y.-T.}\ \bibnamefont {Cheng}}, \bibinfo {author} {\bibfnamefont {C.-H.}\ \bibnamefont {Chien}}, \bibinfo {author} {\bibfnamefont {K.-M.}\ \bibnamefont {Hsieh}}, \bibinfo {author} {\bibfnamefont {Y.-H.}\ \bibnamefont {Huang}}, \bibinfo {author} {\bibfnamefont {P.}~\bibnamefont {Wen}}, \bibinfo {author} {\bibfnamefont {W.-J.}\ \bibnamefont {Lin}}, \bibinfo {author} {\bibfnamefont {Y.}~\bibnamefont {Lu}}, \bibinfo {author} {\bibfnamefont {F.}~\bibnamefont {Aziz}}, \bibinfo {author} {\bibfnamefont {C.-P.}\ \bibnamefont {Lee}}, \bibinfo {author} {\bibfnamefont {K.-T.}\ \bibnamefont {Lin}}, \emph {et~al.},\ }\bibfield  {title} {\bibinfo {title} {Tuning atom-field interaction via phase shaping},\ }\href {https://doi.org/10.1103/PhysRevA.109.023705} {\bibfield  {journal} {\bibinfo  {journal} {Physical Review A}\ }\textbf {\bibinfo {volume} {109}},\ \bibinfo {pages} {023705} (\bibinfo {year} {2024})}\BibitemShut {NoStop}%
\bibitem [{\citenamefont {Wen}\ \emph {et~al.}(2020)\citenamefont {Wen}, \citenamefont {Ivakhnenko}, \citenamefont {Nakonechnyi}, \citenamefont {Suri}, \citenamefont {Lin}, \citenamefont {Lin}, \citenamefont {Chen}, \citenamefont {Shevchenko}, \citenamefont {Nori},\ and\ \citenamefont {Hoi}}]{wen2020landau}%
  \BibitemOpen
  \bibfield  {author} {\bibinfo {author} {\bibfnamefont {P.}~\bibnamefont {Wen}}, \bibinfo {author} {\bibfnamefont {O.}~\bibnamefont {Ivakhnenko}}, \bibinfo {author} {\bibfnamefont {M.}~\bibnamefont {Nakonechnyi}}, \bibinfo {author} {\bibfnamefont {B.}~\bibnamefont {Suri}}, \bibinfo {author} {\bibfnamefont {J.-J.}\ \bibnamefont {Lin}}, \bibinfo {author} {\bibfnamefont {W.-J.}\ \bibnamefont {Lin}}, \bibinfo {author} {\bibfnamefont {J.}~\bibnamefont {Chen}}, \bibinfo {author} {\bibfnamefont {S.}~\bibnamefont {Shevchenko}}, \bibinfo {author} {\bibfnamefont {F.}~\bibnamefont {Nori}},\ and\ \bibinfo {author} {\bibfnamefont {I.-C.}\ \bibnamefont {Hoi}},\ }\bibfield  {title} {\bibinfo {title} {Landau-{Z}ener-{S}t{\"u}ckelberg-{M}ajorana interferometry of a superconducting qubit in front of a mirror},\ }\href {https://doi.org/10.1103/PhysRevB.102.075448} {\bibfield  {journal} {\bibinfo  {journal} {Physical Review B}\ }\textbf {\bibinfo {volume} {102}},\ \bibinfo {pages} {075448} (\bibinfo {year} {2020})}\BibitemShut
  {NoStop}%
\bibitem [{\citenamefont {Liul}\ \emph {et~al.}(2023)\citenamefont {Liul}, \citenamefont {Chien}, \citenamefont {Chen}, \citenamefont {Wen}, \citenamefont {Chen}, \citenamefont {Lin}, \citenamefont {Shevchenko}, \citenamefont {Nori},\ and\ \citenamefont {Hoi}}]{liul2023coherent}%
  \BibitemOpen
  \bibfield  {author} {\bibinfo {author} {\bibfnamefont {M.}~\bibnamefont {Liul}}, \bibinfo {author} {\bibfnamefont {C.-H.}\ \bibnamefont {Chien}}, \bibinfo {author} {\bibfnamefont {C.-Y.}\ \bibnamefont {Chen}}, \bibinfo {author} {\bibfnamefont {P.}~\bibnamefont {Wen}}, \bibinfo {author} {\bibfnamefont {J.}~\bibnamefont {Chen}}, \bibinfo {author} {\bibfnamefont {Y.-H.}\ \bibnamefont {Lin}}, \bibinfo {author} {\bibfnamefont {S.}~\bibnamefont {Shevchenko}}, \bibinfo {author} {\bibfnamefont {F.}~\bibnamefont {Nori}},\ and\ \bibinfo {author} {\bibfnamefont {I.-C.}\ \bibnamefont {Hoi}},\ }\bibfield  {title} {\bibinfo {title} {Coherent dynamics of a photon-dressed qubit},\ }\href {https://doi.org/10.1103/PhysRevB.107.195441} {\bibfield  {journal} {\bibinfo  {journal} {Physical Review B}\ }\textbf {\bibinfo {volume} {107}},\ \bibinfo {pages} {195441} (\bibinfo {year} {2023})}\BibitemShut {NoStop}%
\bibitem [{\citenamefont {Cheng}\ \emph {et~al.}(2025)\citenamefont {Cheng}, \citenamefont {Hsieh}, \citenamefont {Wu}, \citenamefont {Niu}, \citenamefont {Aziz}, \citenamefont {Huang}, \citenamefont {Wen}, \citenamefont {Lin}, \citenamefont {Lin}, \citenamefont {Chen} \emph {et~al.}}]{cheng2025group}%
  \BibitemOpen
  \bibfield  {author} {\bibinfo {author} {\bibfnamefont {Y.-T.}\ \bibnamefont {Cheng}}, \bibinfo {author} {\bibfnamefont {K.-M.}\ \bibnamefont {Hsieh}}, \bibinfo {author} {\bibfnamefont {B.-Y.}\ \bibnamefont {Wu}}, \bibinfo {author} {\bibfnamefont {Z.}~\bibnamefont {Niu}}, \bibinfo {author} {\bibfnamefont {F.}~\bibnamefont {Aziz}}, \bibinfo {author} {\bibfnamefont {Y.-H.}\ \bibnamefont {Huang}}, \bibinfo {author} {\bibfnamefont {P.}~\bibnamefont {Wen}}, \bibinfo {author} {\bibfnamefont {K.-T.}\ \bibnamefont {Lin}}, \bibinfo {author} {\bibfnamefont {Y.-H.}\ \bibnamefont {Lin}}, \bibinfo {author} {\bibfnamefont {J.}~\bibnamefont {Chen}}, \emph {et~al.},\ }\bibfield  {title} {\bibinfo {title} {Group delay controlled by the decoherence of a single artificial atom},\ }\href {https://doi.org/10.1103/fkzb-fxv4} {\bibfield  {journal} {\bibinfo  {journal} {Physical Review Letters}\ }\textbf {\bibinfo {volume} {135}},\ \bibinfo {pages} {073601} (\bibinfo {year} {2025})}\BibitemShut {NoStop}%
\bibitem [{\citenamefont {Bienfait}\ \emph {et~al.}(2019)\citenamefont {Bienfait}, \citenamefont {Satzinger}, \citenamefont {Zhong}, \citenamefont {Chang}, \citenamefont {Chou}, \citenamefont {Conner}, \citenamefont {Dumur}, \citenamefont {Grebel}, \citenamefont {Peairs}, \citenamefont {Povey},\ and\ \citenamefont {Cleland}}]{bienfait2019phonon}%
  \BibitemOpen
  \bibfield  {author} {\bibinfo {author} {\bibfnamefont {A.}~\bibnamefont {Bienfait}}, \bibinfo {author} {\bibfnamefont {K.~J.}\ \bibnamefont {Satzinger}}, \bibinfo {author} {\bibfnamefont {Y.~P.}\ \bibnamefont {Zhong}}, \bibinfo {author} {\bibfnamefont {H.-S.}\ \bibnamefont {Chang}}, \bibinfo {author} {\bibfnamefont {M.-H.}\ \bibnamefont {Chou}}, \bibinfo {author} {\bibfnamefont {C.~R.}\ \bibnamefont {Conner}}, \bibinfo {author} {\bibfnamefont {E.}~\bibnamefont {Dumur}}, \bibinfo {author} {\bibfnamefont {J.}~\bibnamefont {Grebel}}, \bibinfo {author} {\bibfnamefont {G.~A.}\ \bibnamefont {Peairs}}, \bibinfo {author} {\bibfnamefont {R.~G.}\ \bibnamefont {Povey}},\ and\ \bibinfo {author} {\bibfnamefont {A.~N.}\ \bibnamefont {Cleland}},\ }\bibfield  {title} {\bibinfo {title} {Phonon-mediated quantum state transfer and remote qubit entanglement},\ }\href {https://doi.org/10.1126/science.aaw8415} {\bibfield  {journal} {\bibinfo  {journal} {Science}\ }\textbf {\bibinfo {volume} {364}},\ \bibinfo {pages} {368}
  (\bibinfo {year} {2019})}\BibitemShut {NoStop}%
\bibitem [{\citenamefont {Zhang}\ \emph {et~al.}(2022{\natexlab{b}})\citenamefont {Zhang}, \citenamefont {Li}, \citenamefont {Zhang}, \citenamefont {Yuan}, \citenamefont {Chen}, \citenamefont {Ren}, \citenamefont {Wang}, \citenamefont {Song}, \citenamefont {Wang}, \citenamefont {Wang}, \citenamefont {Zhu}, \citenamefont {Agarwal},\ and\ \citenamefont {Scully}}]{zhang2022synthesizing}%
  \BibitemOpen
  \bibfield  {author} {\bibinfo {author} {\bibfnamefont {K.}~\bibnamefont {Zhang}}, \bibinfo {author} {\bibfnamefont {H.}~\bibnamefont {Li}}, \bibinfo {author} {\bibfnamefont {P.}~\bibnamefont {Zhang}}, \bibinfo {author} {\bibfnamefont {J.}~\bibnamefont {Yuan}}, \bibinfo {author} {\bibfnamefont {J.}~\bibnamefont {Chen}}, \bibinfo {author} {\bibfnamefont {W.}~\bibnamefont {Ren}}, \bibinfo {author} {\bibfnamefont {Z.}~\bibnamefont {Wang}}, \bibinfo {author} {\bibfnamefont {C.}~\bibnamefont {Song}}, \bibinfo {author} {\bibfnamefont {D.-W.}\ \bibnamefont {Wang}}, \bibinfo {author} {\bibfnamefont {H.}~\bibnamefont {Wang}}, \bibinfo {author} {\bibfnamefont {S.}~\bibnamefont {Zhu}}, \bibinfo {author} {\bibfnamefont {G.~S.}\ \bibnamefont {Agarwal}},\ and\ \bibinfo {author} {\bibfnamefont {M.~O.}\ \bibnamefont {Scully}},\ }\bibfield  {title} {\bibinfo {title} {Synthesizing five-body interaction in a superconducting quantum circuit},\ }\href {https://doi.org/10.1103/PhysRevLett.128.190502} {\bibfield  {journal}
  {\bibinfo  {journal} {Phys. Rev. Lett.}\ }\textbf {\bibinfo {volume} {128}},\ \bibinfo {pages} {190502} (\bibinfo {year} {2022}{\natexlab{b}})}\BibitemShut {NoStop}%
\bibitem [{\citenamefont {Song}\ \emph {et~al.}(2025)\citenamefont {Song}, \citenamefont {Yang}, \citenamefont {Liu}, \citenamefont {Zhang}, \citenamefont {Xue}, \citenamefont {Mi}, \citenamefont {Zhang}, \citenamefont {Yan}, \citenamefont {Jin},\ and\ \citenamefont {Yu}}]{song2025realization}%
  \BibitemOpen
  \bibfield  {author} {\bibinfo {author} {\bibfnamefont {J.}~\bibnamefont {Song}}, \bibinfo {author} {\bibfnamefont {S.}~\bibnamefont {Yang}}, \bibinfo {author} {\bibfnamefont {P.}~\bibnamefont {Liu}}, \bibinfo {author} {\bibfnamefont {H.-L.}\ \bibnamefont {Zhang}}, \bibinfo {author} {\bibfnamefont {G.-M.}\ \bibnamefont {Xue}}, \bibinfo {author} {\bibfnamefont {Z.-Y.}\ \bibnamefont {Mi}}, \bibinfo {author} {\bibfnamefont {W.-G.}\ \bibnamefont {Zhang}}, \bibinfo {author} {\bibfnamefont {F.}~\bibnamefont {Yan}}, \bibinfo {author} {\bibfnamefont {Y.-R.}\ \bibnamefont {Jin}},\ and\ \bibinfo {author} {\bibfnamefont {H.-F.}\ \bibnamefont {Yu}},\ }\bibfield  {title} {\bibinfo {title} {Realization of high-fidelity perfect entanglers between remote superconducting quantum processors},\ }\href {https://doi.org/10.1103/npr7-b7kq} {\bibfield  {journal} {\bibinfo  {journal} {Phys. Rev. Lett.}\ }\textbf {\bibinfo {volume} {135}},\ \bibinfo {pages} {050603} (\bibinfo {year} {2025})}\BibitemShut {NoStop}%
\bibitem [{\citenamefont {Saha}\ \emph {et~al.}(2025)\citenamefont {Saha}, \citenamefont {Shalaev}, \citenamefont {O'Reilly}, \citenamefont {Goetting}, \citenamefont {Toh}, \citenamefont {Kalakuntla}, \citenamefont {Yu},\ and\ \citenamefont {Monroe}}]{saha2025high}%
  \BibitemOpen
  \bibfield  {author} {\bibinfo {author} {\bibfnamefont {S.}~\bibnamefont {Saha}}, \bibinfo {author} {\bibfnamefont {M.}~\bibnamefont {Shalaev}}, \bibinfo {author} {\bibfnamefont {J.}~\bibnamefont {O'Reilly}}, \bibinfo {author} {\bibfnamefont {I.}~\bibnamefont {Goetting}}, \bibinfo {author} {\bibfnamefont {G.}~\bibnamefont {Toh}}, \bibinfo {author} {\bibfnamefont {A.}~\bibnamefont {Kalakuntla}}, \bibinfo {author} {\bibfnamefont {Y.}~\bibnamefont {Yu}},\ and\ \bibinfo {author} {\bibfnamefont {C.}~\bibnamefont {Monroe}},\ }\bibfield  {title} {\bibinfo {title} {High-fidelity remote entanglement of trapped atoms mediated by time-bin photons},\ }\href {https://doi.org/https://doi.org/10.1038/s41467-025-57557-4} {\bibfield  {journal} {\bibinfo  {journal} {Nature Communications}\ }\textbf {\bibinfo {volume} {16}},\ \bibinfo {pages} {2533} (\bibinfo {year} {2025})}\BibitemShut {NoStop}%
\bibitem [{\citenamefont {Li}\ \emph {et~al.}(2013)\citenamefont {Li}, \citenamefont {Silveri}, \citenamefont {Kumar}, \citenamefont {Pirkkalainen}, \citenamefont {Veps{\"a}l{\"a}inen}, \citenamefont {Chien}, \citenamefont {Tuorila}, \citenamefont {Sillanp{\"a}{\"a}}, \citenamefont {Hakonen}, \citenamefont {Thuneberg} \emph {et~al.}}]{li2013motional}%
  \BibitemOpen
  \bibfield  {author} {\bibinfo {author} {\bibfnamefont {J.}~\bibnamefont {Li}}, \bibinfo {author} {\bibfnamefont {M.}~\bibnamefont {Silveri}}, \bibinfo {author} {\bibfnamefont {K.}~\bibnamefont {Kumar}}, \bibinfo {author} {\bibfnamefont {J.-M.}\ \bibnamefont {Pirkkalainen}}, \bibinfo {author} {\bibfnamefont {A.}~\bibnamefont {Veps{\"a}l{\"a}inen}}, \bibinfo {author} {\bibfnamefont {W.}~\bibnamefont {Chien}}, \bibinfo {author} {\bibfnamefont {J.}~\bibnamefont {Tuorila}}, \bibinfo {author} {\bibfnamefont {M.}~\bibnamefont {Sillanp{\"a}{\"a}}}, \bibinfo {author} {\bibfnamefont {P.}~\bibnamefont {Hakonen}}, \bibinfo {author} {\bibfnamefont {E.}~\bibnamefont {Thuneberg}}, \emph {et~al.},\ }\bibfield  {title} {\bibinfo {title} {Motional averaging in a superconducting qubit},\ }\href {https://doi.org/10.1038/ncomms2383} {\bibfield  {journal} {\bibinfo  {journal} {Nature communications}\ }\textbf {\bibinfo {volume} {4}},\ \bibinfo {pages} {1420} (\bibinfo {year} {2013})}\BibitemShut {NoStop}%
\bibitem [{\citenamefont {Grossmann}\ \emph {et~al.}(1991)\citenamefont {Grossmann}, \citenamefont {Dittrich}, \citenamefont {Jung},\ and\ \citenamefont {H\"{a}nggi}}]{grossmann1991coherent}%
  \BibitemOpen
  \bibfield  {author} {\bibinfo {author} {\bibfnamefont {F.}~\bibnamefont {Grossmann}}, \bibinfo {author} {\bibfnamefont {T.}~\bibnamefont {Dittrich}}, \bibinfo {author} {\bibfnamefont {P.}~\bibnamefont {Jung}},\ and\ \bibinfo {author} {\bibfnamefont {P.}~\bibnamefont {H\"{a}nggi}},\ }\bibfield  {title} {\bibinfo {title} {Coherent destruction of tunneling},\ }\href {https://doi.org/10.1103/PhysRevLett.67.516} {\bibfield  {journal} {\bibinfo  {journal} {Physical review letters}\ }\textbf {\bibinfo {volume} {67}},\ \bibinfo {pages} {516} (\bibinfo {year} {1991})}\BibitemShut {NoStop}%
\bibitem [{\citenamefont {Ashhab}\ \emph {et~al.}(2007)\citenamefont {Ashhab}, \citenamefont {Johansson}, \citenamefont {Zagoskin},\ and\ \citenamefont {Nori}}]{ashhab2007two}%
  \BibitemOpen
  \bibfield  {author} {\bibinfo {author} {\bibfnamefont {S.}~\bibnamefont {Ashhab}}, \bibinfo {author} {\bibfnamefont {J.}~\bibnamefont {Johansson}}, \bibinfo {author} {\bibfnamefont {A.}~\bibnamefont {Zagoskin}},\ and\ \bibinfo {author} {\bibfnamefont {F.}~\bibnamefont {Nori}},\ }\bibfield  {title} {\bibinfo {title} {Two-level systems driven by large-amplitude fields},\ }\href {https://doi.org/10.1103/PhysRevA.75.063414} {\bibfield  {journal} {\bibinfo  {journal} {Physical Review A-Atomic, Molecular, and Optical Physics}\ }\textbf {\bibinfo {volume} {75}},\ \bibinfo {pages} {063414} (\bibinfo {year} {2007})}\BibitemShut {NoStop}%
\bibitem [{\citenamefont {Brehm}\ \emph {et~al.}(2021)\citenamefont {Brehm}, \citenamefont {Poddubny}, \citenamefont {Stehli}, \citenamefont {Wolz}, \citenamefont {Rotzinger},\ and\ \citenamefont {Ustinov}}]{brehm2021waveguide}%
  \BibitemOpen
  \bibfield  {author} {\bibinfo {author} {\bibfnamefont {J.~D.}\ \bibnamefont {Brehm}}, \bibinfo {author} {\bibfnamefont {A.~N.}\ \bibnamefont {Poddubny}}, \bibinfo {author} {\bibfnamefont {A.}~\bibnamefont {Stehli}}, \bibinfo {author} {\bibfnamefont {T.}~\bibnamefont {Wolz}}, \bibinfo {author} {\bibfnamefont {H.}~\bibnamefont {Rotzinger}},\ and\ \bibinfo {author} {\bibfnamefont {A.~V.}\ \bibnamefont {Ustinov}},\ }\bibfield  {title} {\bibinfo {title} {Waveguide bandgap engineering with an array of superconducting qubits},\ }\href {https://doi.org/10.1038/s41535-021-00310-z} {\bibfield  {journal} {\bibinfo  {journal} {npj Quantum Materials}\ }\textbf {\bibinfo {volume} {6}},\ \bibinfo {pages} {10} (\bibinfo {year} {2021})}\BibitemShut {NoStop}%
\bibitem [{\citenamefont {Bello}\ \emph {et~al.}(2022)\citenamefont {Bello}, \citenamefont {Platero},\ and\ \citenamefont {Gonz{\'a}lez-Tudela}}]{bello2022spin}%
  \BibitemOpen
  \bibfield  {author} {\bibinfo {author} {\bibfnamefont {M.}~\bibnamefont {Bello}}, \bibinfo {author} {\bibfnamefont {G.}~\bibnamefont {Platero}},\ and\ \bibinfo {author} {\bibfnamefont {A.}~\bibnamefont {Gonz{\'a}lez-Tudela}},\ }\bibfield  {title} {\bibinfo {title} {Spin many-body phases in standard-and topological-waveguide {QED} simulators},\ }\href {https://doi.org/10.1103/PRXQuantum.3.010336} {\bibfield  {journal} {\bibinfo  {journal} {PRX Quantum}\ }\textbf {\bibinfo {volume} {3}},\ \bibinfo {pages} {010336} (\bibinfo {year} {2022})}\BibitemShut {NoStop}%
\bibitem [{\citenamefont {Zhao}\ \emph {et~al.}(2022)\citenamefont {Zhao}, \citenamefont {Ge}, \citenamefont {Xiang}, \citenamefont {Xue}, \citenamefont {Yan}, \citenamefont {Wang}, \citenamefont {Wang}, \citenamefont {Xu}, \citenamefont {Su}, \citenamefont {Yang} \emph {et~al.}}]{zhao2022probing}%
  \BibitemOpen
  \bibfield  {author} {\bibinfo {author} {\bibfnamefont {S.}~\bibnamefont {Zhao}}, \bibinfo {author} {\bibfnamefont {Z.-Y.}\ \bibnamefont {Ge}}, \bibinfo {author} {\bibfnamefont {Z.}~\bibnamefont {Xiang}}, \bibinfo {author} {\bibfnamefont {G.}~\bibnamefont {Xue}}, \bibinfo {author} {\bibfnamefont {H.}~\bibnamefont {Yan}}, \bibinfo {author} {\bibfnamefont {Z.}~\bibnamefont {Wang}}, \bibinfo {author} {\bibfnamefont {Z.}~\bibnamefont {Wang}}, \bibinfo {author} {\bibfnamefont {H.}~\bibnamefont {Xu}}, \bibinfo {author} {\bibfnamefont {F.}~\bibnamefont {Su}}, \bibinfo {author} {\bibfnamefont {Z.}~\bibnamefont {Yang}}, \emph {et~al.},\ }\bibfield  {title} {\bibinfo {title} {Probing operator spreading via floquet engineering in a superconducting circuit},\ }\href {https://doi.org/10.1103/PhysRevLett.129.160602} {\bibfield  {journal} {\bibinfo  {journal} {Physical Review Letters}\ }\textbf {\bibinfo {volume} {129}},\ \bibinfo {pages} {160602} (\bibinfo {year} {2022})}\BibitemShut {NoStop}%
\bibitem [{\citenamefont {Salath{\'e}}\ \emph {et~al.}(2015)\citenamefont {Salath{\'e}}, \citenamefont {Mondal}, \citenamefont {Oppliger}, \citenamefont {Heinsoo}, \citenamefont {Kurpiers}, \citenamefont {Poto{\v{c}}nik}, \citenamefont {Mezzacapo}, \citenamefont {Las~Heras}, \citenamefont {Lamata}, \citenamefont {Solano} \emph {et~al.}}]{salathe2015digital}%
  \BibitemOpen
  \bibfield  {author} {\bibinfo {author} {\bibfnamefont {Y.}~\bibnamefont {Salath{\'e}}}, \bibinfo {author} {\bibfnamefont {M.}~\bibnamefont {Mondal}}, \bibinfo {author} {\bibfnamefont {M.}~\bibnamefont {Oppliger}}, \bibinfo {author} {\bibfnamefont {J.}~\bibnamefont {Heinsoo}}, \bibinfo {author} {\bibfnamefont {P.}~\bibnamefont {Kurpiers}}, \bibinfo {author} {\bibfnamefont {A.}~\bibnamefont {Poto{\v{c}}nik}}, \bibinfo {author} {\bibfnamefont {A.}~\bibnamefont {Mezzacapo}}, \bibinfo {author} {\bibfnamefont {U.}~\bibnamefont {Las~Heras}}, \bibinfo {author} {\bibfnamefont {L.}~\bibnamefont {Lamata}}, \bibinfo {author} {\bibfnamefont {E.}~\bibnamefont {Solano}}, \emph {et~al.},\ }\bibfield  {title} {\bibinfo {title} {Digital quantum simulation of spin models with circuit quantum electrodynamics},\ }\href {https://doi.org/10.1103/PhysRevX.5.021027} {\bibfield  {journal} {\bibinfo  {journal} {Physical Review X}\ }\textbf {\bibinfo {volume} {5}},\ \bibinfo {pages} {021027} (\bibinfo {year} {2015})}\BibitemShut
  {NoStop}%
\bibitem [{\citenamefont {Shankar}\ \emph {et~al.}(2013)\citenamefont {Shankar}, \citenamefont {Hatridge}, \citenamefont {Leghtas}, \citenamefont {Sliwa}, \citenamefont {Narla}, \citenamefont {Vool}, \citenamefont {Girvin}, \citenamefont {Frunzio}, \citenamefont {Mirrahimi},\ and\ \citenamefont {Devoret}}]{shankar2013autonomously}%
  \BibitemOpen
  \bibfield  {author} {\bibinfo {author} {\bibfnamefont {S.}~\bibnamefont {Shankar}}, \bibinfo {author} {\bibfnamefont {M.}~\bibnamefont {Hatridge}}, \bibinfo {author} {\bibfnamefont {Z.}~\bibnamefont {Leghtas}}, \bibinfo {author} {\bibfnamefont {K.}~\bibnamefont {Sliwa}}, \bibinfo {author} {\bibfnamefont {A.}~\bibnamefont {Narla}}, \bibinfo {author} {\bibfnamefont {U.}~\bibnamefont {Vool}}, \bibinfo {author} {\bibfnamefont {S.~M.}\ \bibnamefont {Girvin}}, \bibinfo {author} {\bibfnamefont {L.}~\bibnamefont {Frunzio}}, \bibinfo {author} {\bibfnamefont {M.}~\bibnamefont {Mirrahimi}},\ and\ \bibinfo {author} {\bibfnamefont {M.~H.}\ \bibnamefont {Devoret}},\ }\bibfield  {title} {\bibinfo {title} {Autonomously stabilized entanglement between two superconducting quantum bits},\ }\href {https://doi.org/10.1038/nature12802} {\bibfield  {journal} {\bibinfo  {journal} {Nature}\ }\textbf {\bibinfo {volume} {504}},\ \bibinfo {pages} {419} (\bibinfo {year} {2013})}\BibitemShut {NoStop}%
\bibitem [{\citenamefont {Kurpiers}\ \emph {et~al.}(2018)\citenamefont {Kurpiers}, \citenamefont {Magnard}, \citenamefont {Walter}, \citenamefont {Royer}, \citenamefont {Pechal}, \citenamefont {Heinsoo}, \citenamefont {Salath{\'e}}, \citenamefont {Akin}, \citenamefont {Storz}, \citenamefont {Besse} \emph {et~al.}}]{kurpiers2018deterministic}%
  \BibitemOpen
  \bibfield  {author} {\bibinfo {author} {\bibfnamefont {P.}~\bibnamefont {Kurpiers}}, \bibinfo {author} {\bibfnamefont {P.}~\bibnamefont {Magnard}}, \bibinfo {author} {\bibfnamefont {T.}~\bibnamefont {Walter}}, \bibinfo {author} {\bibfnamefont {B.}~\bibnamefont {Royer}}, \bibinfo {author} {\bibfnamefont {M.}~\bibnamefont {Pechal}}, \bibinfo {author} {\bibfnamefont {J.}~\bibnamefont {Heinsoo}}, \bibinfo {author} {\bibfnamefont {Y.}~\bibnamefont {Salath{\'e}}}, \bibinfo {author} {\bibfnamefont {A.}~\bibnamefont {Akin}}, \bibinfo {author} {\bibfnamefont {S.}~\bibnamefont {Storz}}, \bibinfo {author} {\bibfnamefont {J.-C.}\ \bibnamefont {Besse}}, \emph {et~al.},\ }\bibfield  {title} {\bibinfo {title} {Deterministic quantum state transfer and remote entanglement using microwave photons},\ }\href {https://doi.org/10.1038/s41586-018-0195-y} {\bibfield  {journal} {\bibinfo  {journal} {Nature}\ }\textbf {\bibinfo {volume} {558}},\ \bibinfo {pages} {264} (\bibinfo {year} {2018})}\BibitemShut {NoStop}%
\bibitem [{\citenamefont {Shah}\ \emph {et~al.}(2024)\citenamefont {Shah}, \citenamefont {Yang}, \citenamefont {Joshi},\ and\ \citenamefont {Mirhosseini}}]{shah2024stabilizing}%
  \BibitemOpen
  \bibfield  {author} {\bibinfo {author} {\bibfnamefont {P.~S.}\ \bibnamefont {Shah}}, \bibinfo {author} {\bibfnamefont {F.}~\bibnamefont {Yang}}, \bibinfo {author} {\bibfnamefont {C.}~\bibnamefont {Joshi}},\ and\ \bibinfo {author} {\bibfnamefont {M.}~\bibnamefont {Mirhosseini}},\ }\bibfield  {title} {\bibinfo {title} {Stabilizing remote entanglement via waveguide dissipation},\ }\href {https://doi.org/10.1103/PRXQuantum.5.030346} {\bibfield  {journal} {\bibinfo  {journal} {PRX Quantum}\ }\textbf {\bibinfo {volume} {5}},\ \bibinfo {pages} {030346} (\bibinfo {year} {2024})}\BibitemShut {NoStop}%
\bibitem [{\citenamefont {Kannan}\ \emph {et~al.}(2020)\citenamefont {Kannan}, \citenamefont {Campbell}, \citenamefont {Vasconcelos}, \citenamefont {Winik}, \citenamefont {Kim}, \citenamefont {Kjaergaard}, \citenamefont {Krantz}, \citenamefont {Melville}, \citenamefont {Niedzielski}, \citenamefont {Yoder} \emph {et~al.}}]{kannan2020generating}%
  \BibitemOpen
  \bibfield  {author} {\bibinfo {author} {\bibfnamefont {B.}~\bibnamefont {Kannan}}, \bibinfo {author} {\bibfnamefont {D.~L.}\ \bibnamefont {Campbell}}, \bibinfo {author} {\bibfnamefont {F.}~\bibnamefont {Vasconcelos}}, \bibinfo {author} {\bibfnamefont {R.}~\bibnamefont {Winik}}, \bibinfo {author} {\bibfnamefont {D.}~\bibnamefont {Kim}}, \bibinfo {author} {\bibfnamefont {M.}~\bibnamefont {Kjaergaard}}, \bibinfo {author} {\bibfnamefont {P.}~\bibnamefont {Krantz}}, \bibinfo {author} {\bibfnamefont {A.}~\bibnamefont {Melville}}, \bibinfo {author} {\bibfnamefont {B.~M.}\ \bibnamefont {Niedzielski}}, \bibinfo {author} {\bibfnamefont {J.}~\bibnamefont {Yoder}}, \emph {et~al.},\ }\bibfield  {title} {\bibinfo {title} {Generating spatially entangled itinerant photons with waveguide quantum electrodynamics},\ }\href {https://doi.org/10.1126/sciadv.abb8780} {\bibfield  {journal} {\bibinfo  {journal} {Science advances}\ }\textbf {\bibinfo {volume} {6}},\ \bibinfo {pages} {eabb8780} (\bibinfo {year} {2020})}\BibitemShut
  {NoStop}%
\bibitem [{\citenamefont {Krastanov}\ \emph {et~al.}(2021)\citenamefont {Krastanov}, \citenamefont {Raniwala}, \citenamefont {Holzgrafe}, \citenamefont {Jacobs}, \citenamefont {Lon{\v{c}}ar}, \citenamefont {Reagor},\ and\ \citenamefont {Englund}}]{krastanov2021optically}%
  \BibitemOpen
  \bibfield  {author} {\bibinfo {author} {\bibfnamefont {S.}~\bibnamefont {Krastanov}}, \bibinfo {author} {\bibfnamefont {H.}~\bibnamefont {Raniwala}}, \bibinfo {author} {\bibfnamefont {J.}~\bibnamefont {Holzgrafe}}, \bibinfo {author} {\bibfnamefont {K.}~\bibnamefont {Jacobs}}, \bibinfo {author} {\bibfnamefont {M.}~\bibnamefont {Lon{\v{c}}ar}}, \bibinfo {author} {\bibfnamefont {M.~J.}\ \bibnamefont {Reagor}},\ and\ \bibinfo {author} {\bibfnamefont {D.~R.}\ \bibnamefont {Englund}},\ }\bibfield  {title} {\bibinfo {title} {Optically heralded entanglement of superconducting systems in quantum networks},\ }\href {https://doi.org/10.1103/PhysRevLett.127.040503} {\bibfield  {journal} {\bibinfo  {journal} {Physical Review Letters}\ }\textbf {\bibinfo {volume} {127}},\ \bibinfo {pages} {040503} (\bibinfo {year} {2021})}\BibitemShut {NoStop}%
\bibitem [{\citenamefont {Diehl}\ \emph {et~al.}(2010)\citenamefont {Diehl}, \citenamefont {Tomadin}, \citenamefont {Micheli}, \citenamefont {Fazio},\ and\ \citenamefont {Zoller}}]{diehl2010dynamical}%
  \BibitemOpen
  \bibfield  {author} {\bibinfo {author} {\bibfnamefont {S.}~\bibnamefont {Diehl}}, \bibinfo {author} {\bibfnamefont {A.}~\bibnamefont {Tomadin}}, \bibinfo {author} {\bibfnamefont {A.}~\bibnamefont {Micheli}}, \bibinfo {author} {\bibfnamefont {R.}~\bibnamefont {Fazio}},\ and\ \bibinfo {author} {\bibfnamefont {P.}~\bibnamefont {Zoller}},\ }\bibfield  {title} {\bibinfo {title} {Dynamical phase transitions and instabilities in open atomic many-body systems},\ }\href {https://doi.org/10.1103/PhysRevLett.105.015702} {\bibfield  {journal} {\bibinfo  {journal} {Physical review letters}\ }\textbf {\bibinfo {volume} {105}},\ \bibinfo {pages} {015702} (\bibinfo {year} {2010})}\BibitemShut {NoStop}%
\bibitem [{\citenamefont {Fitzpatrick}\ \emph {et~al.}(2017)\citenamefont {Fitzpatrick}, \citenamefont {Sundaresan}, \citenamefont {Li}, \citenamefont {Koch},\ and\ \citenamefont {Houck}}]{fitzpatrick2017observation}%
  \BibitemOpen
  \bibfield  {author} {\bibinfo {author} {\bibfnamefont {M.}~\bibnamefont {Fitzpatrick}}, \bibinfo {author} {\bibfnamefont {N.~M.}\ \bibnamefont {Sundaresan}}, \bibinfo {author} {\bibfnamefont {A.~C.}\ \bibnamefont {Li}}, \bibinfo {author} {\bibfnamefont {J.}~\bibnamefont {Koch}},\ and\ \bibinfo {author} {\bibfnamefont {A.~A.}\ \bibnamefont {Houck}},\ }\bibfield  {title} {\bibinfo {title} {Observation of a dissipative phase transition in a one-dimensional circuit {QED} lattice},\ }\href {https://doi.org/10.1103/PhysRevX.7.011016} {\bibfield  {journal} {\bibinfo  {journal} {Physical Review X}\ }\textbf {\bibinfo {volume} {7}},\ \bibinfo {pages} {011016} (\bibinfo {year} {2017})}\BibitemShut {NoStop}%
\bibitem [{\citenamefont {Diehl}\ \emph {et~al.}(2011)\citenamefont {Diehl}, \citenamefont {Rico}, \citenamefont {Baranov},\ and\ \citenamefont {Zoller}}]{diehl2011topology}%
  \BibitemOpen
  \bibfield  {author} {\bibinfo {author} {\bibfnamefont {S.}~\bibnamefont {Diehl}}, \bibinfo {author} {\bibfnamefont {E.}~\bibnamefont {Rico}}, \bibinfo {author} {\bibfnamefont {M.~A.}\ \bibnamefont {Baranov}},\ and\ \bibinfo {author} {\bibfnamefont {P.}~\bibnamefont {Zoller}},\ }\bibfield  {title} {\bibinfo {title} {Topology by dissipation in atomic quantum wires},\ }\href {https://doi.org/10.1038/nphys2106} {\bibfield  {journal} {\bibinfo  {journal} {Nature physics}\ }\textbf {\bibinfo {volume} {7}},\ \bibinfo {pages} {971} (\bibinfo {year} {2011})}\BibitemShut {NoStop}%
\bibitem [{\citenamefont {Bergholtz}\ \emph {et~al.}(2021)\citenamefont {Bergholtz}, \citenamefont {Budich},\ and\ \citenamefont {Kunst}}]{bergholtz2021exceptional}%
  \BibitemOpen
  \bibfield  {author} {\bibinfo {author} {\bibfnamefont {E.~J.}\ \bibnamefont {Bergholtz}}, \bibinfo {author} {\bibfnamefont {J.~C.}\ \bibnamefont {Budich}},\ and\ \bibinfo {author} {\bibfnamefont {F.~K.}\ \bibnamefont {Kunst}},\ }\bibfield  {title} {\bibinfo {title} {Exceptional topology of non-hermitian systems},\ }\href {https://doi.org/10.1103/RevModPhys.93.015005} {\bibfield  {journal} {\bibinfo  {journal} {Reviews of Modern Physics}\ }\textbf {\bibinfo {volume} {93}},\ \bibinfo {pages} {015005} (\bibinfo {year} {2021})}\BibitemShut {NoStop}%
\bibitem [{\citenamefont {Bello}\ \emph {et~al.}(2019)\citenamefont {Bello}, \citenamefont {Platero}, \citenamefont {Cirac},\ and\ \citenamefont {Gonz{\'a}lez-Tudela}}]{bello2019unconventional}%
  \BibitemOpen
  \bibfield  {author} {\bibinfo {author} {\bibfnamefont {M.}~\bibnamefont {Bello}}, \bibinfo {author} {\bibfnamefont {G.}~\bibnamefont {Platero}}, \bibinfo {author} {\bibfnamefont {J.~I.}\ \bibnamefont {Cirac}},\ and\ \bibinfo {author} {\bibfnamefont {A.}~\bibnamefont {Gonz{\'a}lez-Tudela}},\ }\bibfield  {title} {\bibinfo {title} {Unconventional quantum optics in topological waveguide qed},\ }\href {https://doi.org/10.1126/sciadv.aaw0297} {\bibfield  {journal} {\bibinfo  {journal} {Science advances}\ }\textbf {\bibinfo {volume} {5}},\ \bibinfo {pages} {eaaw0297} (\bibinfo {year} {2019})}\BibitemShut {NoStop}%
\bibitem [{\citenamefont {Koch}\ \emph {et~al.}(2007)\citenamefont {Koch}, \citenamefont {Yu}, \citenamefont {Gambetta}, \citenamefont {Houck}, \citenamefont {Schuster}, \citenamefont {Majer}, \citenamefont {Blais}, \citenamefont {Devoret}, \citenamefont {Girvin},\ and\ \citenamefont {Schoelkopf}}]{koch2007charge}%
  \BibitemOpen
  \bibfield  {author} {\bibinfo {author} {\bibfnamefont {J.}~\bibnamefont {Koch}}, \bibinfo {author} {\bibfnamefont {T.~M.}\ \bibnamefont {Yu}}, \bibinfo {author} {\bibfnamefont {J.}~\bibnamefont {Gambetta}}, \bibinfo {author} {\bibfnamefont {A.~A.}\ \bibnamefont {Houck}}, \bibinfo {author} {\bibfnamefont {D.~I.}\ \bibnamefont {Schuster}}, \bibinfo {author} {\bibfnamefont {J.}~\bibnamefont {Majer}}, \bibinfo {author} {\bibfnamefont {A.}~\bibnamefont {Blais}}, \bibinfo {author} {\bibfnamefont {M.~H.}\ \bibnamefont {Devoret}}, \bibinfo {author} {\bibfnamefont {S.~M.}\ \bibnamefont {Girvin}},\ and\ \bibinfo {author} {\bibfnamefont {R.~J.}\ \bibnamefont {Schoelkopf}},\ }\bibfield  {title} {\bibinfo {title} {Charge-insensitive qubit design derived from the cooper pair box},\ }\href {https://doi.org/10.1103/PhysRevA.76.042319} {\bibfield  {journal} {\bibinfo  {journal} {Physical Review A-Atomic, Molecular, and Optical Physics}\ }\textbf {\bibinfo {volume} {76}},\ \bibinfo {pages} {042319} (\bibinfo {year}
  {2007})}\BibitemShut {NoStop}%
\bibitem [{\citenamefont {Probst}\ \emph {et~al.}(2015)\citenamefont {Probst}, \citenamefont {Song}, \citenamefont {Bushev}, \citenamefont {Ustinov},\ and\ \citenamefont {Weides}}]{probst2015efficient}%
  \BibitemOpen
  \bibfield  {author} {\bibinfo {author} {\bibfnamefont {S.}~\bibnamefont {Probst}}, \bibinfo {author} {\bibfnamefont {F.}~\bibnamefont {Song}}, \bibinfo {author} {\bibfnamefont {P.~A.}\ \bibnamefont {Bushev}}, \bibinfo {author} {\bibfnamefont {A.~V.}\ \bibnamefont {Ustinov}},\ and\ \bibinfo {author} {\bibfnamefont {M.}~\bibnamefont {Weides}},\ }\bibfield  {title} {\bibinfo {title} {Efficient and robust analysis of complex scattering data under noise in microwave resonators},\ }\href {https://doi.org/10.1063/1.4907935} {\bibfield  {journal} {\bibinfo  {journal} {Review of Scientific Instruments}\ }\textbf {\bibinfo {volume} {86}} (\bibinfo {year} {2015})}\BibitemShut {NoStop}%
\bibitem [{\citenamefont {Lu}\ \emph {et~al.}(2021)\citenamefont {Lu}, \citenamefont {Bengtsson}, \citenamefont {Burnett}, \citenamefont {Wiegand}, \citenamefont {Suri}, \citenamefont {Krantz}, \citenamefont {Roudsari}, \citenamefont {Kockum}, \citenamefont {Gasparinetti}, \citenamefont {Johansson} \emph {et~al.}}]{lu2021characterizing}%
  \BibitemOpen
  \bibfield  {author} {\bibinfo {author} {\bibfnamefont {Y.}~\bibnamefont {Lu}}, \bibinfo {author} {\bibfnamefont {A.}~\bibnamefont {Bengtsson}}, \bibinfo {author} {\bibfnamefont {J.~J.}\ \bibnamefont {Burnett}}, \bibinfo {author} {\bibfnamefont {E.}~\bibnamefont {Wiegand}}, \bibinfo {author} {\bibfnamefont {B.}~\bibnamefont {Suri}}, \bibinfo {author} {\bibfnamefont {P.}~\bibnamefont {Krantz}}, \bibinfo {author} {\bibfnamefont {A.~F.}\ \bibnamefont {Roudsari}}, \bibinfo {author} {\bibfnamefont {A.~F.}\ \bibnamefont {Kockum}}, \bibinfo {author} {\bibfnamefont {S.}~\bibnamefont {Gasparinetti}}, \bibinfo {author} {\bibfnamefont {G.}~\bibnamefont {Johansson}}, \emph {et~al.},\ }\bibfield  {title} {\bibinfo {title} {Characterizing decoherence rates of a superconducting qubit by direct microwave scattering},\ }\href {https://doi.org/10.1038/s41534-021-00367-5} {\bibfield  {journal} {\bibinfo  {journal} {npj Quantum Information}\ }\textbf {\bibinfo {volume} {7}},\ \bibinfo {pages} {35} (\bibinfo {year}
  {2021})}\BibitemShut {NoStop}%
\end{thebibliography}%
\end{document}


\title{Supplemental Material for \\``Collective Floquet engineering between two distant superconducting artificial atoms"}

\author{Y.-H.~Huang}
\thanks{These authors contributed equally}
\affiliation{Department of Physics, National Tsing Hua University, Hsinchu 30013, Taiwan}

\author{Jiale~Yuan}
\thanks{These authors contributed equally}
\email[e-mail:]{jialeyuan@csu.edu.cn}
\affiliation{School of Physics, Central South University, Changsha 410083, China}
\affiliation{Zhejiang Key Laboratory of Micro-Nano Quantum Chips and Quantum Control, School of Physics, and State Key Laboratory for Extreme Photonics and Instrumentation, Zhejiang University, Hangzhou 310027, China}

\author{Z.~Q.~Niu}
\thanks{These authors contributed equally}
\affiliation{State Key Laboratory of Materials for Integrated Circuits, Shanghai Institute of Microsystem and Information Technology, Chinese Academy of Sciences, Shanghai 200050, China}

\author{Y.-S.~Tsai}
\affiliation{Department of Physics, National Tsing Hua University, Hsinchu 30013, Taiwan}

\author{K.-C.~Yu}
\affiliation{Department of Physics, National Tsing Hua University, Hsinchu 30013, Taiwan}

\author{P.-H.~Hsu}
\affiliation{Department of Physics, National Tsing Hua University, Hsinchu 30013, Taiwan}

\author{K.-M.~Hsieh}
\affiliation{Department of Physics, City University of Hong Kong, Kowloon, Hong Kong SAR 999077, China}

\author{Haw-Tyng Huang}
\affiliation{Department of Materials Science and Engineering, National Taiwan University, Taipei, 10617, Taiwan}

\author{J.~C.~Chen}
\affiliation{Department of Physics, National Tsing Hua University, Hsinchu 30013, Taiwan}
\affiliation{Center for Quantum Technology, National Tsing Hua University, Hsinchu 30013, Taiwan}

\author{P.~Y.~Wen}
\affiliation{Department of Physics, National Chung Cheng University, Chiayi 621301, Taiwan}

\author{Y.-H.~Lin}
\affiliation{Department of Physics, National Tsing Hua University, Hsinchu 30013, Taiwan}
\affiliation{Center for Quantum Technology, National Tsing Hua University, Hsinchu 30013, Taiwan}

\author{Z.-R.~Lin}
\email[e-mail:]{zrlin@mail.sim.ac.cn}
\affiliation{State Key Laboratory of Materials for Integrated Circuits, Shanghai Institute of Microsystem and Information Technology, Chinese Academy of Sciences, Shanghai 200050, China}

\author{Da-Wei~Wang}
\affiliation{Zhejiang Key Laboratory of Micro-Nano Quantum Chips and Quantum Control, School of Physics, and State Key Laboratory for Extreme Photonics and Instrumentation, Zhejiang University, Hangzhou 310027, China}

\author{I.-C.~Hoi}
\email[e-mail:]{iochoi@cityu.edu.hk}
\affiliation{Department of Physics, City University of Hong Kong, Kowloon, Hong Kong SAR 999077, China}

\date{\today}

\maketitle

\tableofcontents


\section{Experimental setup}
\begin{figure}
\includegraphics[width=\linewidth]{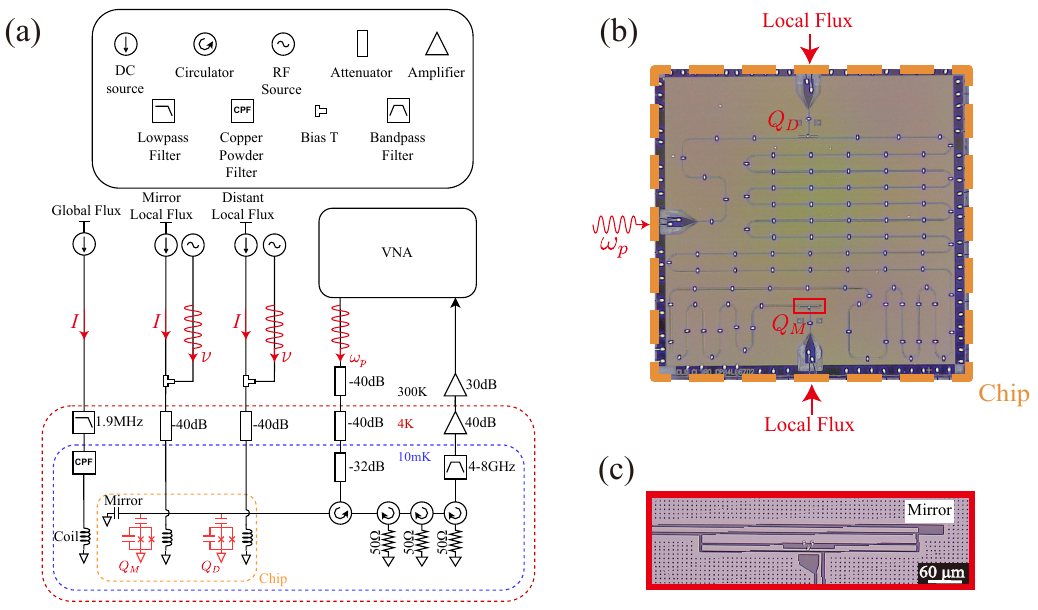}
\caption{
Measurement setup and device images. (a) The experimental setup. The device under test (orange dashed box) is mounted at the base stage of a dilution refrigerator with a base temperature of \unit[10]{mK} (blue dashed box). The reflected signal is measured by VNA, with $\omega_p$ denoting the probe frequency. Each qubit is equipped with a local flux line, where the flux is composed of a DC bias and an oscillating pump at frequency $\nu$, pump power $P_{pump}$, and phase $\theta_{m,d}$, combined via a bias-T. (b) Optical microscope images of the chip [orange dashed box in (a)]. The transmission line is terminated with a capacitive load that functions as a mirror. The mirror qubit ($Q_M$) and the distant qubit ($Q_D$) are transmon-type artificial atoms~\cite{Koch2007} capacitively coupled to the waveguide. $Q_M$ is highlighted by the red box, with an enlarged view provided in (c). $Q_D$ features a design identical to $Q_M$ but is positioned 60 mm from the mirror. (c) High-magnification micrograph of $Q_M$, illustrating the transmon geometry used for both qubits. 
}
\label{fig:exp_setup}
\end{figure}

The experimental setup is illustrated in \figref{fig:exp_setup}, featuring a sample box containing a chip (highlighted by the orange dashed box) with two superconducting artificial atoms capacitively coupled to a 1D semi-infinite transmission line. This assembly is mounted at the base stage of a dilution refrigerator with a base temperature of \unit[10]{mK}. A superconducting magnetic coil, centered on the sample box, provides a global magnetic flux to tune the qubits' transition frequencies. We measure the continuous-wave (CW) reflected coefficient of the two-qubit system using a vector network analyzer (VNA), with $\omega_p$ denoting the probe frequency. In the detection chain, the probe signal undergoes a total effective attenuation of \unit[-131]{dB} before reaching the sample; after interacting with the system, the signal is amplified by an effective gain of \unit[75]{dB} by a two-stage amplification process involving a high-electron-mobility transistor (HEMT) amplifier at \unit[3]{K} and another amplifier at room temperature. These gain and attenuation values are obtained by fitting the power-dependent response of a single transmon qubit at a fixed frequency \cite{cheng2024tuning}; the values deviate slightly within the frequency range we operate (less than \unit[1]{dB}). 

Each qubit is modulated via its local flux line independently, where the flux is composed of a DC bias and an oscillating pump at frequency $\nu$, pump power $P_{pump}$, and phase $\theta_{m,d}$, combined using a bias-T at room temperature, followed by \unit[-40]{dB} of attenuation in the cryogenic environment. Additionally, they are both biased by a global flux via a coil. The bias-T is a microwave component that combines DC and RF sources; the DC port is characterized by inductive input impedance, while the RF port is capacitively coupled.

Optical micrographs of the chip are shown in \figpanel{fig:exp_setup}{b}. The transmission line is a superconducting coplanar waveguide (CPW) terminated with a capacitive load that serves as a mirror. We fabricated two transmon-type artificial atoms~\cite{Koch2007} on the chip, with one placed adjacent to the mirror ($Q_M$, mirror qubit) and the other 60 mm away from the mirror ($Q_D$, distant qubit). The two qubits have identical designs and are both capacitively coupled to the waveguide.

\section{Single-tone spectroscopy}
\label{sec:Singletone_spectroscopy}
\begin{figure}
\includegraphics[width=\linewidth]{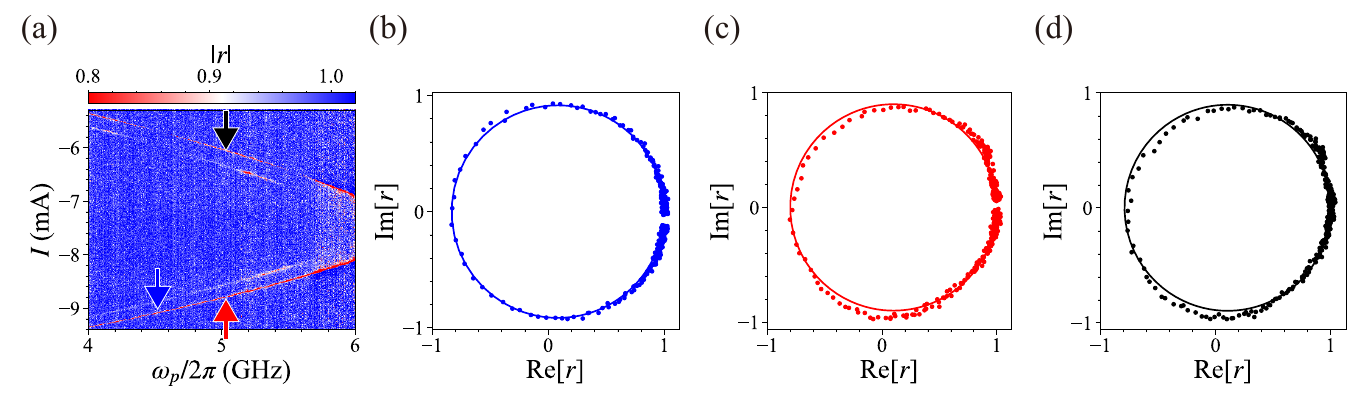}
\caption{
Single-tone spectroscopy and device characterization. (a) Magnitude of reflection coefficient $\left | r\right |$ as a function of $\omega_p$ and bias current $I$ controlling the $\ket{0}-\ket{1}$ transition frequencies $\omega_{10}$ of both qubits through the coil (see \figref{fig:exp_setup}). The red, blue, and black arrows indicate the bias points used to extract the qubits' parameters shown in (b), (c), and (d), respectively. (b)-(d) IQ plots of the complex reflection coefficient $r$ at the bias points indicated in (a). Experimental data (dots) are fitted with theoretical curves (solid lines) using the standard circle-fit method \cite{probst2015efficient, lu2021characterizing}. The extracted qubits' parameters for $Q_M$ and $Q_D$ are summarized in Table \ref{tab:qubit_param}, including $\omega_{10}$, relaxation rate $\Gamma_{10}$, pure dephasing rate $\Gamma^\phi_1$, and total decoherence rate $\gamma_{10}$.
\label{fig:single_tone}
}
\end{figure}

To characterize the qubit parameters, we conducted single-tone spectroscopy. The magnitude of the reflection coefficient $\left|r\right|$ is illustrated in \figpanel{fig:single_tone}{a} as a function of probe frequency $\omega_p$ and bias current $I$. The current $I$ controls the global magnetic flux $\Phi$, which tunes the qubits' transition frequencies $\omega_{10}$. The spectrum reveals the characteristic cosine flux dependence of the transmon $\ket{0}-\ket{1}$ transition:
\begin{equation}
    \hbar \omega_{10} (\Phi) \approx \sqrt{8E_j(\Phi)E_c}-E_c
\end{equation}
The transition energy is determined by the charging energy $E_c = e^2/ 2C_{total}$ and the Josephson energy:
\begin{equation}
    E_j(\Phi) = E_{j,Max}|\cos (\pi \Phi/ \Phi_0)|
\end{equation}
where $\Phi_0 = h/2e$ is the magnetic flux quantum. The Josephson energy can be tuned from its maximum value $E_{j, Max}$ by the external flux $\Phi$ via a magnetic coil or local flux line.
\figpanels{fig:single_tone}{b}{d} depict the reflection coefficient in the IQ plane under weak probing ($\Omega_p\ll\gamma_{10}$), resulting in characteristic circular resonance patterns. \figand{fig:single_tone}{b}{c} correspond to $Q_M$ resonance at $f_1/2\pi=\unit[4.527]{GHz}$ and $f_2/2\pi=\unit[5.031]{GHz}$, respectively, while \figpanel{fig:single_tone}{d} shows $Q_D$ resonance at $f_2$. Experimental data points (dots) are colored to match the corresponding operating frequencies marked in \figpanel{fig:single_tone}{a}; the solid curves represent the theoretical fits \cite{probst2015efficient, lu2021characterizing}, from which we extract the parameters $\omega_{0}$, relaxation rate $\Gamma_{10}$, and decoherence rate $\gamma_{10}$. The extracted parameters are shown in Table \ref{tab:qubit_param}.
We note that in \figand{fig:single_tone}{c}{d}, the experimental data deviate slightly from the theoretical fit; we attribute this to stray cavity resonances visible at $\omega_p/2\pi=\unit[5.01]{GHz}$ and $\omega_p/2\pi=\unit[5.075]{GHz}$ [see $\delta_p/2\pi=\unit[-21]{MHz}$ and $\delta_p/2\pi=\unit[44]{MHz}$ in \figpanel{fig:CLS}{b}]. This resonance induces the asymmetric patterns observed in all measurements performed at $f_2$.

        
            
        


\setlength{\tabcolsep}{15pt}
\begin{table*}
    \centering
    \scalebox{0.9}{
        \begin{tabular}{| c | c | c | c | c | c | c | c |}
            \hline
            Qubit &
            $E_{c}/h$ &
            $E_{j}/h$ &
            $E_{j}/E_{c}$ &
            $\omega_{10}/2\pi$ &
            $\Gamma_{10}/2\pi$ &
            $\Gamma^\phi_1/2\pi$ &
            $\gamma_{10}/2\pi$ \\
            
            \hline
            - &
            [MHz] &
            [GHz] &
            [--] &
            [GHz] &
            [MHz] &
            [MHz] &
            [MHz] \\
            
            \hline
            \textcolor{blue}{$\rightarrow$}\, $Q_M(\textit{f}_1)$ &
            369.1 &
            29.14 &
            78.96 &
            4.527 &
            12.13 &
            0.469 &
            6.535 \\
        
            \hline
            \textcolor{red}{$\rightarrow$}\, $Q_M(\textit{f}_2)$ &
            369.1 &
            29.14 &
            78.96 &
            5.031 &
            17.53 &
            0.913 &
            9.677 \\
            
            \hline
            $\rightarrow$\, $Q_D(\textit{f}_2)$ &
            380.8 &
            25.26 &
            66.34 &
            5.031 &
            17.64 &
            0.928 &
            9.745 \\
            
            \hline
        \end{tabular}
    }

\caption{
Extracted parameters of $Q_D$ and $Q_M$. The pure dephasing rate $\Gamma^\phi_1$ is calculated by the relation $\gamma_{10} = \Gamma_{10}/2 + \Gamma^\phi$.
}
\label{tab:qubit_param}
\end{table*}

\section{Static qubit-qubit interaction}
\label{sec:static}
\begin{figure}
\includegraphics[width=\linewidth]{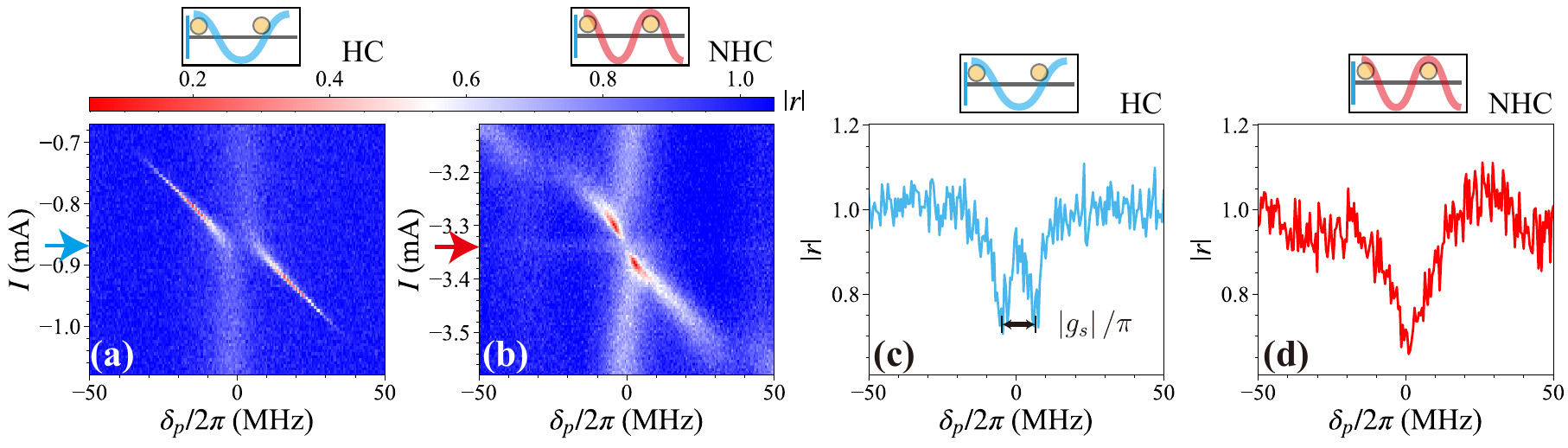}
\caption{
Static qubit-qubit interaction. (a), (b) Magnitude of reflection coefficient $\left | r \right|$ as a function of $\delta_p$ and bias current $I$ applied to $Q_D$, measured without any flux pump with $\omega_m$ fixed at operating frequencies $f_1$ (a), and $f_2$ (b). The bias current $I$ tunes $\omega_{d}$ through resonance with $\omega_m$. At $f_1$, $Q_D$ resides at a node of standing wave and cannot decay into the waveguide; the qubit-qubit interaction is therefore mediated by virtual photons, giving rise to an HC interaction with strength $\left | g_s \right|$ that manifests as an avoided crossing \cite{wen2019large, lin2019scalable, van2013photon, bao2024level}. At $f_2$, both qubits are at antinodes and couple directly to the waveguide, and the interaction is instead NHC, mediated by real photons, manifesting as a crossing together with an enhanced
collective linewidth \cite{van2013photon, bao2024level}. The blue and red arrows indicate the on-resonance bias points at $f_1$ and $f_2$, respectively, which correspond to the unpumped qubit-qubit interaction conditions used as the starting points for the Floquet engineering experiments in the main text. In (b), additional stray resonances appear near $\delta_p/2\pi=\unit[-21]{MHz}$ and $\delta_p/2\pi=\unit[44]{MHz}$, which we attribute to imperfections in the experimental setup. (c), (d) Linecuts of $\left | r \right|$ taken at the bias-current positions marked by the arrows in (a) and (b), respectively. The blue curve in (c) resolves two separated resonances corresponding to the hybridized collective modes in the HC configuration. The red curve in (d) shows no coherent splitting; instead, the response is governed by collective radiative decay in the NHC configuration.
}
\label{fig:CLS}

\end{figure}

The static qubit-qubit interaction is characterized at both operating frequencies, $f_1$ and $f_2$, by sweeping the bias current $I$ through the local flux line of $Q_D$ \cite{wen2019large}, which tunes $\omega_{d}$ independently through resonance with $\omega_m$. The results are shown in \figref{fig:CLS}.
At $f_1$, $Q_D$ resides at a node of standing wave, so the non-Hermitian coupling (NHC) channel is completely suppressed and the qubit-qubit interaction reduces to a Hermitian coupling (HC) $g_{s}\in \mathbb{R}$. This manifests as an avoided crossing in [\figand{fig:CLS}{a}{c}], from which the HC strength $\left | g_s \right|$ can be directly extracted. At $f_2$, $Q_D$ resides at an antinode, reversing the situation so that the HC channel vanishes and the interaction becomes non-Hermitian with $g_{s}\in i\mathbb{R}$, mediated by real photons. This manifests as a level crossing together with an enhanced collective linewidth, as observed in [\figand{fig:CLS}{b}{d}]. The on-resonance bias points at $f_1$ and $f_2$ are indicated by the blue and red arrows, respectively, and serve as the static starting configurations upon which the Floquet drive is applied in all subsequent experiments. To make the distinction between the two static configurations more explicit, we take linecuts of the reflection spectra at the bias-current positions indicated by the arrows, as shown in \figpanel{fig:CLS}{c} and \figpanel{fig:CLS}{d}. In the HC configuration at $f_1$ [\figpanel{fig:CLS}{c}], the two minima in $|r|$ directly reveal the normal-mode splitting $2\left | g_s \right|$ of the hybridized qubit states. In contrast, at $f_2$ [\figpanel{fig:CLS}{d}], the two collective modes are degenerate in frequency, so no coherent splitting is resolved in the line cut; their collective nature is instead encoded in the NHC response through different radiative linewidths.

\section{Theoretical model}
\label{section:model}
\subsection{Numerical simulation of the reflection coefficient}
In this section, we derive the effective model and the equations for numerically solving the reflection spectrum.
Our setup consists of two qubits coupled through a one-dimensional waveguide terminated by a large capacitor, which is equivalent to placing a mirror at the end of the waveguide.
The mirror creates a standing wave in the sample. A mirror qubit $Q_M$, placed next to the mirror, always sits at an antinode and is maximally coupled to the waveguide. A distant qubit is placed away from the mirror, with a tunable effective distance ($L/\lambda$).
The Hamiltonian of our system is $H = H_S+H_B+H_{int}$. The atomic part is $H_S =  \hbar \omega_m \sigma_m^+\sigma_m^- + \hbar \omega_d \sigma_d^+\sigma_d^-$, where $\sigma_j^+$ and $\sigma_j^-$ are the raising and lowering operators of the qubit, $\omega_j$ is the transition frequency, and $j=m$, $d$ denote the mirror and distant qubits, respectively.
The field part is $H_B = \int_0^\infty\hbar\omega a_\omega^\dagger a_\omega d\omega$, where $a_\omega^\dagger$ ($a_\omega$) creates (annihilates) a photon of frequency $\omega$ in the waveguide, and the operators satisfy the commutation relation $[a_\omega', a_\omega^\dagger]=\delta(\omega-\omega')$. The interaction part is $H_{int}=i\sum_{j=m,d}\int_0^\infty d\omega \hbar g_{\omega,j}\cos(k_\omega x_j) a_\omega \sigma_j^+ + \text{H.c.}$, where $g_{\omega,j}$ is the coupling strength between the photon mode at frequency $\omega$ and the $j$-th qubit, and the mode function $\cos(k_\omega x)$ is due to the antinode mirror at $x=0$. 

Following the standard procedure of tracing out the photonic degrees of freedom and applying the Born-Markov approximation, we obtain the master equation ($\hbar=1$)~\cite{lin2019scalable} 
\begin{equation}
\begin{aligned}
\frac{d\rho}{dt}=&-i\sum_{j=m,d}\omega_j [\sigma_j^+\sigma_j^-,\rho]\\
&-i\sum_{j,k=m,d}(\tilde{\gamma}_{jk}^+-i\gamma_{jk}^-)[\sigma_j^+\sigma_k^-,\rho]\\
&-i\sum_{j=m,d}\Omega_p^j\cos(k_px_j)[e^{-i\omega_p t}\sigma_j^+ + e^{i\omega_p t}\sigma_j^-,\rho]\\
&+\sum_{j,k=m,d}(\gamma_{jk}^+ +i\tilde{\gamma}_{jk}^-)\mathcal{L}_{jk}[\rho]\\
&+\sum_{j=m,d}\Gamma_j^\phi \mathcal{L}_{j}^\phi[\rho].
\label{eq:master}
\end{aligned}
\end{equation}
The third term corresponds to a continuous-wave probe field incident from the other end
of the waveguide with frequency $\omega_p$, wavenumber $k_p$, and Rabi frequency $\Omega_p^j\cos(k_px_j)$ experienced by the $j$-th qubit.
The superoperator $\mathcal{L}_{jk}[\rho]=2\sigma_j^-\rho \sigma_k^+ - \sigma_j^+\sigma_k^-\rho - \rho\sigma_j^+\sigma_k^-$ describes individual and collective dissipation processes, and $\mathcal{L}_{j}^\phi[\rho]=2\sigma_j^{ee}\rho \sigma_j^{ee} - \sigma_j^{ee}\rho - \rho\sigma_j^{ee}$ with $\sigma_j^{ee}=|e\rangle_j\langle e|_j$ describes individual pure dephasing processes. The inter-qubit
interaction, obtained by summing all contributions from the waveguide modes, is given by $\gamma_{ij}^\pm = (\gamma_{ij}\pm \gamma_{ji})/2$ and $\tilde{\gamma}_{ij}^\pm = (\tilde{\gamma}_{ij}\pm \tilde{\gamma}_{ji})/2$ with
\begin{equation}
\begin{aligned}
\gamma_{ij} = \frac{\gamma_{ij}^0}{2}\left[\cos k_j(x_i+x_j) + \cos k_j|x_i-x_j|\right],\\
\tilde{\gamma}_{ij} = \frac{\gamma_{ij}^0}{2}\left[\sin k_j(x_i+x_j) + \sin k_j|x_i-x_j|\right],
\label{eq:gamma}
\end{aligned}
\end{equation}
where $\gamma_{ij}^0 = \sqrt{\Gamma_i(\omega_j)\Gamma_j(\omega_j)}$ with $\Gamma_i = \pi g_i^2(\omega_j)$ being the bare relaxation rate. For identical qubits, $\gamma_{ij}^-$ and $\tilde{\gamma}_{ij}^-$ vanish, which we assume hereafter.

The reflection coefficient is governed by the steady-state density matrix elements $\rho_{mg}$ and $\rho_{dg}$. From the master equation \eqref{eq:master}, we obtain
\begin{equation}
\begin{aligned}
    \frac{\partial \rho_{mg}}{\partial t}= &-i(\Omega_p^m e^{-i\omega_pt}+(\omega_m + \tilde{\gamma}_{mm}^+)\rho_{mg} + \tilde{\gamma}_{md}^+\rho_{dg})-(\gamma_{mm}^++\Gamma_{m}^\phi) \rho_{mg} - \gamma_{md}^+\rho_{dg},\\
\frac{\partial \rho_{dg}}{\partial t}= &-i(\Omega_p^d e^{-i\omega_pt} +(\omega_d + \tilde{\gamma}_{dd}^+)\rho_{dg}+ \tilde{\gamma}_{dm}^+ \rho_{mg} )-  (\gamma_{dd}^+ +\Gamma_{d}^\phi)\rho_{dg}- \gamma_{dm}^+\rho_{mg}.
\end{aligned}
\end{equation}
Under sinusoidal frequency modulation, the qubit transition frequency is $\omega_j = f_i + \Delta_j\cos(\nu t + \theta_j)$. Here only the phase difference $\theta = \theta_d-\theta_m$ matters, and without loss of generality, we assume $\theta_d=\theta$ and $\theta_m=0$.
Setting $\rho_{jg} = \tilde{\rho}_j e^{-i\omega_p t}$, we obtain
\begin{equation}
\begin{aligned}
 \frac{\partial\tilde{\rho}_m}{\partial t}-i\delta_p\tilde{\rho}_m= &-i(\Omega_p^m +[\Delta_m \cos(\nu t )+\tilde{\gamma}_{mm}^+]\tilde{\rho}_m+ \tilde{\gamma}_{md}^+\tilde{\rho}_d)-(\gamma_{mm}^++\Gamma_{m}^\phi) \tilde{\rho}_m - \gamma_{md}^+\tilde{\rho}_d,\\
 \frac{\partial \tilde{\rho}_{d}}{\partial t}-i\delta_p \tilde{\rho}_d= &-i(\Omega_p^d +[\Delta_d \cos(\nu t +\theta)+\tilde{\gamma}_{dd}^+]\tilde{\rho}_d+ \tilde{\gamma}_{dm}^+\tilde{\rho}_m)-(\gamma_{dd}^++\Gamma_{d}^\phi) \tilde{\rho}_d - \gamma_{dm}^+\tilde{\rho}_m,
\end{aligned}
\end{equation}
where $\delta_p = \omega_p - f_i$ is the probe field detuning.
By expanding the density matrix elements into their Fourier components $\tilde{\rho}_j = \sum_n e^{in\nu t}\rho_j^{[n]}$ and setting $\partial_t\rho_j^{[n]}=0$ for a steady state, we obtain
\begin{equation}
\begin{aligned}
in\nu\rho_m^{[n]} -i\delta_p \rho_m^{[n]} = &-i(\Omega_p^m\delta_{\text{kr}}[n]+\frac{\Delta_m }{2}\rho_m^{[n-1]}+\frac{\Delta_m }{2}\rho_m^{[n+1]}+\tilde{\gamma}_{mm}^+\rho_m^{[n]}+ \tilde{\gamma}_{md}^+\rho_{d}^{[n]} )-(\gamma_{mm}^++\Gamma_{m}^\phi)\rho_{m}^{[n]} -\gamma_{md}^+\rho_{d}^{[n]}, \\
in\nu\rho_d^{[n]} -i\delta_p \rho_d^{[n]}= &-i(\Omega_p^d \delta_{\text{kr}}[n]  +\frac{\Delta_d }{2}e^{i\theta}\rho_d^{[n-1]}+\frac{\Delta_d }{2}e^{-i\theta}\rho_d^{[n+1]} + \tilde{\gamma}_{dd}^+\rho_d^{[n]}  +\tilde{\gamma}_{dm}^+\rho_{m}^{[n]}) - (\gamma_{dd}^+ +\Gamma_{d}^\phi) \rho_{d}^{[n]} -\gamma_{dm}^+\rho_{m}^{[n]},
\end{aligned}
\end{equation}
where $\delta_{\text{kr}}[\cdot]$ is the Kronecker symbol. Simplifying the above equations give the steady-state equations
\begin{equation}
\begin{aligned}
(in\nu -i\delta_p+i\tilde{\gamma}_{mm}^+ +\gamma_{mm}^+ +\Gamma_{m}^\phi) \rho_m^{[n]} +\rho_m^{[n-1]}(i\frac{\Delta_m }{2} )+\rho_m^{[n+1]}(i\frac{\Delta_m }{2} ) +(i \tilde{\gamma}_{md}^+ + \gamma_{md}^+)\rho_{d}^{[n]} = -i\Omega_p^m\delta_{\text{kr}}[n],\\
(in\nu -i\delta_p+i\tilde{\gamma}_{dd}^++\gamma_{dd}^+ +  \Gamma_{d}^\phi) \rho_d^{[n]} + \rho_d^{[n-1]}(i\frac{\Delta_d }{2} e^{i\theta})+ \rho_d^{[n+1]}(i\frac{\Delta_d }{2} e^{-i\theta}) +(i \tilde{\gamma}_{dm}^+ + \gamma_{dm}^+)\rho_{m}^{[n]} = -i\Omega_p^d\delta_{\text{kr}}[n].
\label{eq:solve_steady_state}
\end{aligned}
\end{equation}
In the simulation, we numerically solve \eqref{eq:solve_steady_state} to obtain $\rho_m^{[0]}$ and $\rho_d^{[0]}$, which govern the reflection spectrum.

When the distant qubit is placed at a node, we have $x_m=0$ and $x_d = \pi/(2k_0)$, where $k_0 = \omega_{10}/v$ and $v$ is the speed of light in the waveguide. From \eqref{eq:gamma} we obtain $\gamma_{md}^+ = \gamma_{dm}^+ = \gamma_{dd}^+ = 0$, $\gamma_{mm}^+ = \gamma_{mm}^0$, $\tilde{\gamma}_{mm}^+ = \tilde{\gamma}_{dd}^+ = 0$, and $\tilde{\gamma}_{md}^+ = \tilde{\gamma}_{dm}^+ = \gamma_{md}^0$. Since the probe field only couples to the mirror qubit, the reflection coefficient is solely determined by $\rho_{m}^{[0]}$ as $1 - i\eta_c\rho_{m}^{[0]}/\Omega_p^m$, where $\eta_c$ is a constant related to the capacitors and Josephson energies of the sample. 

When the distant qubit is placed at an antinode, we have $x_m=0$ and $x_d = 2\pi/k_0$, yielding $\tilde{\gamma}_{mm}^+ = \tilde{\gamma}_{dd}^+ = \tilde{\gamma}_{md}^+ = \tilde{\gamma}_{dm}^+=0$, $\gamma_{mm}^+ = \gamma_{mm}^0$, $\gamma_{dd}^+ = \gamma_{dd}^0$, $\gamma_{md}^+ =\gamma_{dm}^+ = \gamma_{dm}^0$. The probe field couples to both qubits, and the reflection coefficient is $1 - i\eta_c(\rho_{m}^{[0]}/\Omega_p^m + \rho_{d}^{[0]}/\Omega_p^d)$. These produce the numerical simulations presented in \figref{fig:Fig2_sup}, and \figref{fig:Fig2_sup_-32dBm}-\figref{fig:Fig4_sup}.


\subsection{Derivation of the effective Hamiltonian}
It is instructive to derive the effective Floquet Hamiltonian that helps us to understand the underlying physics. In the frame rotating at frequency $f_i$, we write the effective Hamiltonian by neglecting the loss terms in the master equation as
\begin{equation}
\begin{aligned}
H = \Delta_d\cos(\nu t+\theta_d)\frac{\sigma_{d}^{z}}{2} + \Delta_m\cos(\nu t+\theta_m)\frac{\sigma_{m}^{z}}{2}+ g_{s}(\sigma_m^+\sigma_d^- + \sigma_d^+\sigma_m^-),
\end{aligned}
\end{equation}
where $g_{s}$ denotes the effective static coupling strength between the qubits, and $\sigma_j^z$ denotes the $z$-component of the Pauli matrices of the $j$-th qubit. When the distant qubit is at a node, we have $g_{s}=\gamma_{md}^0$, corresponding to the HC configuration. When the distant qubit is at an antinode, we have $g_{s}=-i\gamma_{md}^0$, rendering the coupling term non-Hermitian.

We first consider the case where a single qubit, say without loss of generality the distant qubit, is driven. The Hamiltonian reduces to (set $\theta_d=0$)
$H^{(1)} = (\Delta_d/2)\cos(\nu t)\sigma_{d}^{z} + g_{s}(\sigma_m^+\sigma_d^- + \sigma_d^+\sigma_m^-)$.
In the interaction picture, we have 
\begin{equation}
\begin{aligned}
H_I^{(1)} = &g_{s}(e^{-if\sin\nu t}\sigma_m^+\sigma_d^- + e^{if\sin\nu t}\sigma_d^+\sigma_m^-)\\
= &g_{s}\sum_n \left[J_n(-f)e^{in\nu t}\sigma_m^+\sigma_d^- + J_n(f)e^{in\nu t}\sigma_d^+\sigma_m^-\right],
\label{eqefip}
\end{aligned}
\end{equation}
where $f = \Delta_d/\nu$ is the ratio between the modulation amplitude and frequency. The last equality is derived using the expansion $e^{if\sin\phi} = \sum_{n=-\infty}^{\infty}J_n(f) e^{in\phi}$, where $J_n$ is the $n$-th Bessel function of the first kind.
This suggests that a Bessel function $J_n(f)$ dresses the coupling strength between Floquet states differing from $n$ Floquet bands in both HC and NHC configurations.

We then consider the case where both qubits are driven with the same modulation amplitude $\Delta_m = \Delta_d$.
The Hamiltonian in the interaction picture is (set $\theta_d=-\theta_m=\theta/2$)
\begin{equation}
\begin{aligned}
H_I^{(2)} =& g_{s}(e^{i[f\sin(\nu t-\theta/2) - f\sin(\nu t+\theta/2)]}\sigma_m^+\sigma_d^- + e^{-i[f\sin(\nu t-\theta/2) - f\sin(\nu t+\theta/2)]}\sigma_d^+\sigma_m^-)\\
=&g_{s}\sum_n \left\{J_n[2f\sin(\theta/2)](-i)^ne^{in\nu t}\sigma_m^+\sigma_d^- + J_n[2f\sin(\theta/2)]i^ne^{in\nu t}\sigma_d^+\sigma_m^-\right\},
\label{eqefi}
\end{aligned}
\end{equation}
This yields a $\theta$-dependent coefficient $J_n[2f\sin(\theta/2)]$ dressing the coupling between Floquet states differing from $n$ Floquet bands in both HC and NHC cases. 

Solving these Hamiltonians gives the eigenvalue spectra in the lower panels of \figref{fig:Fig2_sup}, and \figref{fig:Fig2_sup_-32dBm}-\figref{fig:Fig4_sup}. In the HC configuration, where the distant qubit is placed at a node, the eigenvalue spectra fit well with the peak positions of the reflection spectra. In the NHC configuration, the real parts of the eigenvalue spectra follow $E_n=n\nu$, while the imaginary parts determine the linewidths of the bands in the reflection spectra.



\section{Floquet-engineered qubit-qubit interaction}
\label{Supplementary fig}
\begin{figure}[h]
\includegraphics[width=\linewidth]{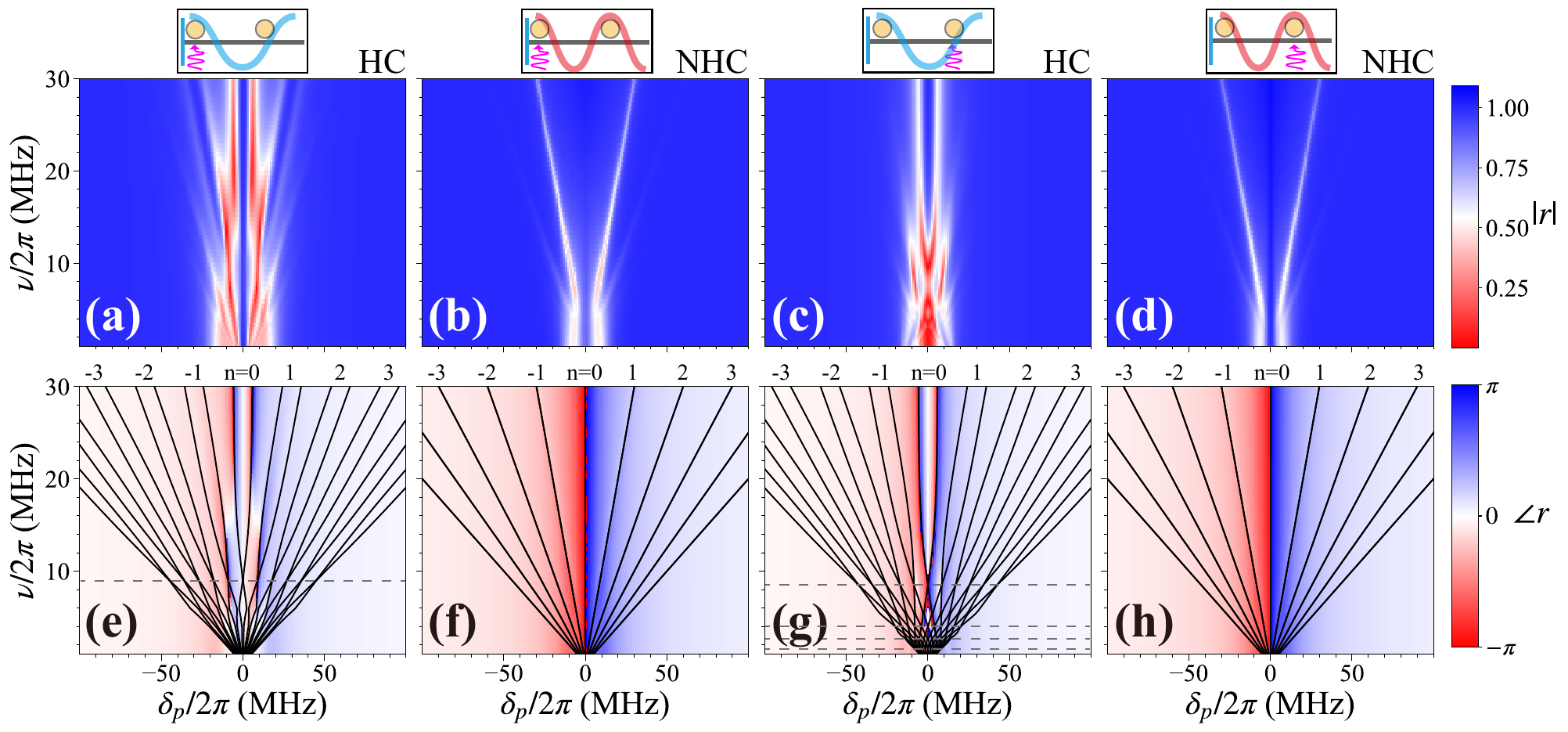}
\caption{
Numerical simulations of the Floquet-engineered qubit-qubit coupling as a function of $\nu$ at $P_{pump}\approx\unit[-80]{dBm}$, corresponding to Figs. 2(a)-(d) in the main text. The coupling mechanism is illustrated in Fig. 1(d) of the main text. (a)-(d) magnitude of reflection coefficient $\left|r\right|$ and (e)-(h) phase response $\angle r$ as a function of $\delta_p$ and $\nu$. A flux pump is applied to $Q_M$ in (a), (b), (e), and (f), and to $Q_D$ in (c), (d), (g), and (h), as illustrated by the pump-condition schematics above each panel. Panels (a), (c), (e), (g) correspond to $f_1$, where both qubits are biased at the HC configuration; panels (b), (d), (f), (h) correspond to $f_2$, where both qubits are biased at the NHC configuration. The black solid curves overlaid on the phase response [(e)-(h)] are eigenvalues calculated from \eqref{eqefip}, showing the Floquet quasienergy dispersion $\delta_p=n\nu$ and confirming the correspondence between the spectral branch structure and the single-qubit-driven Floquet ladder. The horizontal dashed lines in (e) and (g) mark the $\nu$ values at which the corresponding $g_n$ vanish. The numbers above the lower panels [(e)-(h)] label the Floquet band index $n$ of the corresponding quasienergy branches, with $n=0$ denoting the central band and $n=\pm1, \pm2, \pm3$ denoting the drive-assisted sideband channels.
}
\label{fig:Fig2_sup}
\end{figure}

\begin{figure}[h]
\includegraphics[width=\linewidth]{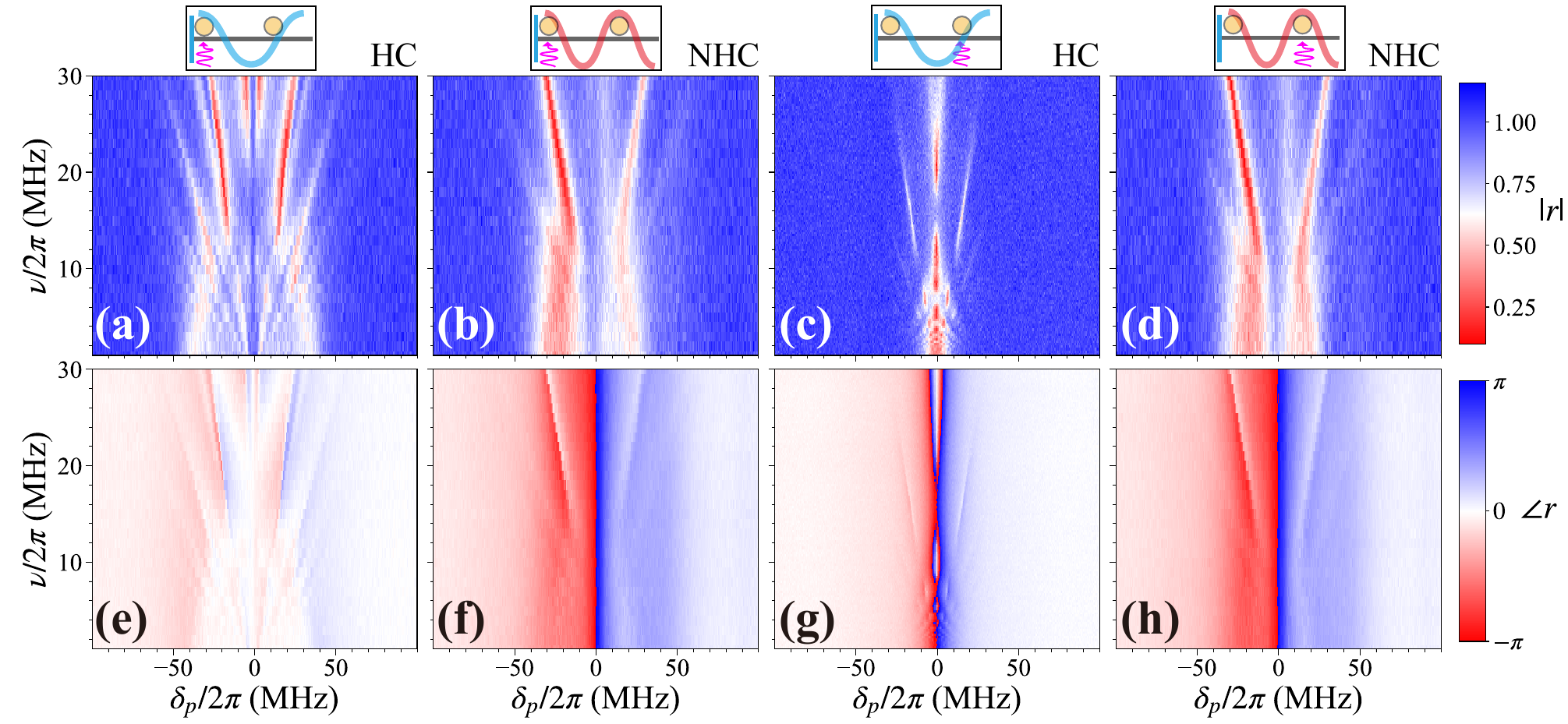}
\caption{
Floquet-engineered qubit-qubit coupling as a function of $\nu$ at an elevated pump power of  $P_{pump}\approx\unit[-72]{dBm}$, measured under the same configuration as Figs. 2(a)-(d) in the main text but with a larger $f$. (a)-(d) Magnitude reflection coefficient $\left|r\right|$ and (e)-(h) phase response $\angle r$ as a function of $\delta_p$ and $\nu$. A flux pump is applied to $Q_M$ in (a), (b), (e), and (f), and to $Q_D$ in (c), (d), (g), and (h), as illustrated by the pump-condition schematics in the upper left corner of each panel. Panels (a), (c), (e), (g) correspond to $f_1$, where both qubits are biased at the HC configuration; panels (b), (d), (f), (h) correspond to $f_2$, where both qubits are biased at the NHC configuration. Compared to the results in the main text, increased $P_{pump}$ activates more Floquet sidebands, producing a denser multi-branch fan at $f_1$ and broader NHC spectral response at $f_2$.
}
\label{fig:Fig2_sup_-32dBm_expt}
\end{figure}

\begin{figure}[h]
\includegraphics[width=\linewidth]{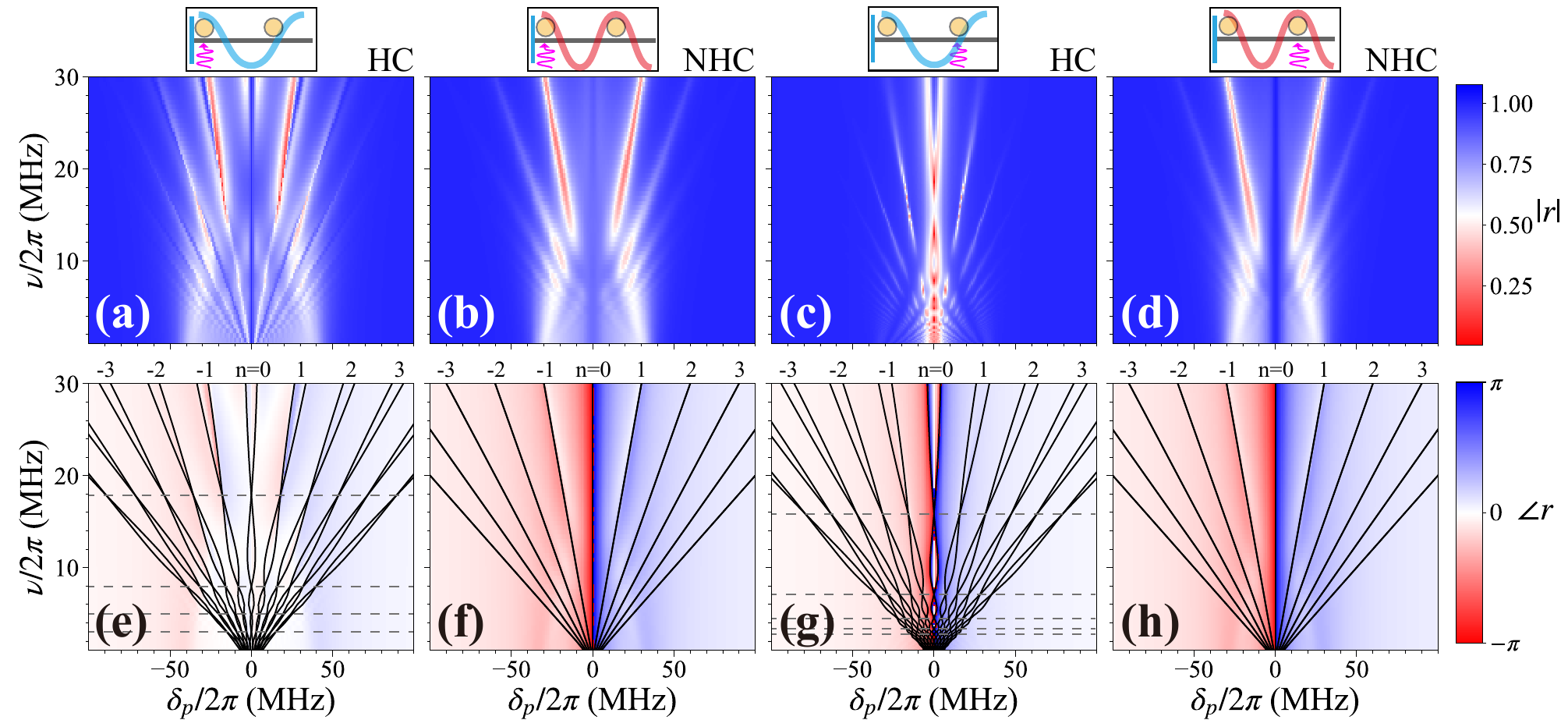}
\caption{
Numerical simulations of the Floquet-engineered qubit-qubit coupling as a function of $\nu$ at $P_{pump}\approx\unit[-72]{dBm}$, corresponding to the experimental data in \figref{fig:Fig2_sup_-32dBm_expt}. The coupling mechanism is illustrated in Fig. 1(d) of the main text. (a)-(d) Magnitude of reflection coefficient $\left|r\right|$ and (e)-(h) phase response $\angle r$ as a function of $\delta_p$ and $\nu$. A flux pump is applied to $Q_M$ in (a), (b), (e), and (f), and to $Q_D$ in (c), (d), (g), and (h), as illustrated by the pump-condition schematics above each panel. Panels (a), (c), (e), (g) correspond to $f_1$, where both qubits are biased at the HC configuration; panels (b), (d), (f), (h) correspond to $f_2$, where both qubits are biased at the NHC configuration. The black solid curves overlaid on the phase response [(e)-(h)] are eigenvalues calculated from \eqref{eqefip}, reproducing the denser branch structure at this higher drive parameter $f$ and showing good agreement with the experimental data in \figref{fig:Fig2_sup_-32dBm_expt}. The horizontal dashed lines in (e) and (g) mark the $\nu$ values at which the corresponding $g_n$ vanish. The numbers above the lower panels [(e)-(h)] indicate the Floquet band index $n$ associated with each calculated quasienergy branch; compared with \figref{fig:Fig2_sup}, the larger pump power gives stronger higher-order sideband weights, making the $n=\pm 2,\pm 3$ branches more visible.
}
\label{fig:Fig2_sup_-32dBm}
\end{figure}

\figref{fig:Fig2_sup} shows the numerical simulations corresponding to Figs. 2(a)-(d) of the main text, presenting all four pumping configurations for both operating frequencies $f_1$ and $f_2$. \figref{fig:Fig2_sup_-32dBm_expt} shows the experimental data measured at a higher pump power of $P_{pump}\approx\unit[-72]{dBm}$, where the larger drive parameter $f$ activates more Floquet sidebands, producing a denser multi-branch fan structure at $f_1$ and broader NHC linewidth response at $f_2$. \figref{fig:Fig2_sup_-32dBm} shows the corresponding numerical simulations for \figref{fig:Fig2_sup_-32dBm_expt}, confirming good agreement with the experimental observations across all four pumping configurations. In the phase response panels of Figs. \ref{fig:Fig2_sup} and \ref{fig:Fig2_sup_-32dBm}, the black curves overlaid on each panel indicate the positions of the Floquet sideband resonances analytically predicted by \eqref{eqefip}, where each branch satisfies $\delta_p=n\nu$ for integer $n$, showing good agreement with the numerical results across both pumping configurations and both operating frequencies. Meanwhile,  the horizontal dashed lines mark the $\nu$ values at which the corresponding $g_n$ vanish, indicating the zeros of the associated Floquet coupling channels. 

\begin{figure}[h]
\includegraphics[width=\linewidth]{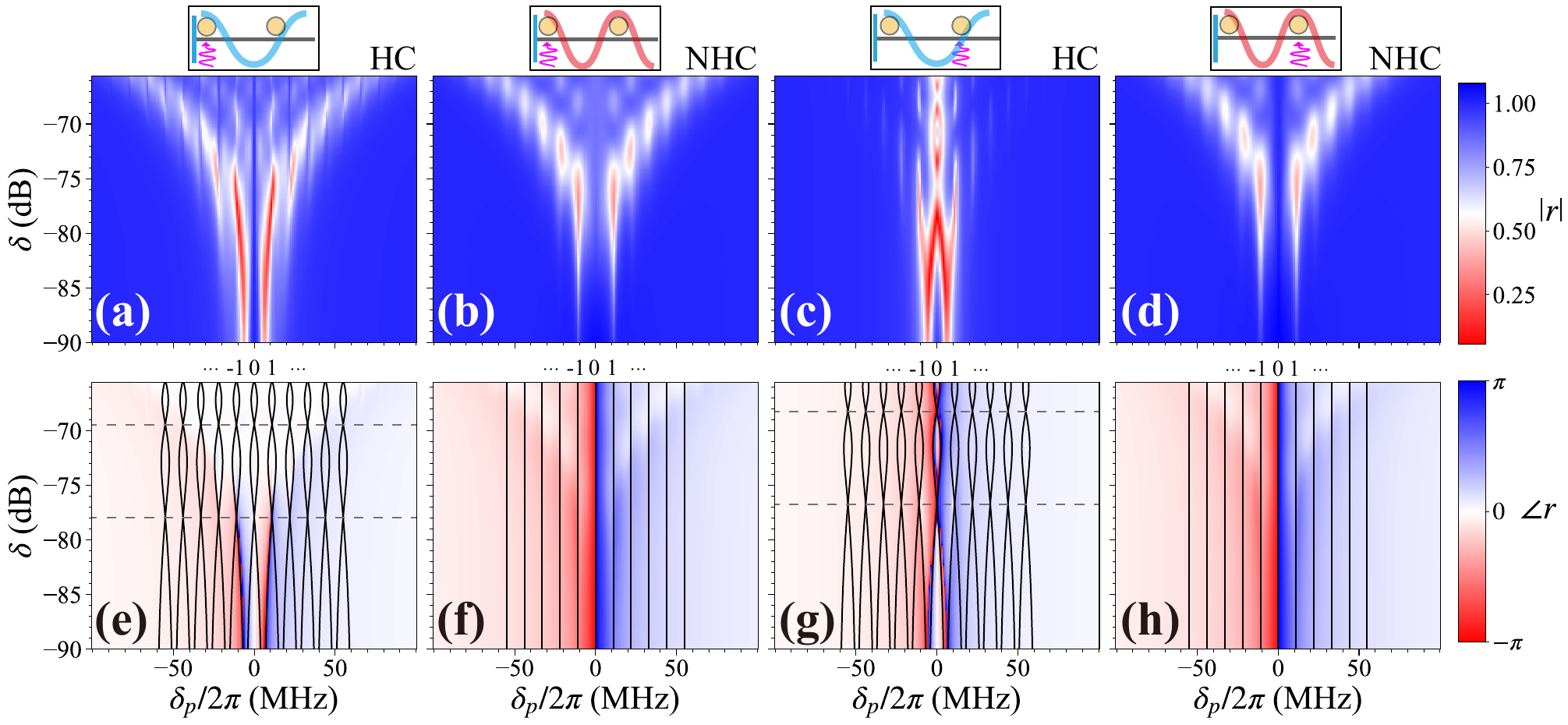}
\caption{
Numerical simulations of the Floquet-engineered qubit-qubit coupling as a function of $\delta=\Delta_i/\Delta_0$, and $i=m$, $d$ for the mirror and distant qubits, respectively. $\Delta_0$ is the modulation amplitude corresponding to the reference pump power used in the experiment. The coupling mechanism is illustrated in Fig. 1(d) of the main text. (a)-(d) Magnitude of reflection coefficient $\left|r\right|$ and (e)-(h) phase response $\angle r$ as a function of $\delta_p$ and $\delta$ at fixed $\nu/2\pi=\unit[11]{MHz}$. A flux pump is applied to $Q_M$ in (a), (b), (e), and (f), and to $Q_D$ in (c), (d), (g), and (h), as illustrated by the pump-condition schematics above each panel. Panels (a), (c), (e), (g) correspond to $f_1$, where both qubits are biased at the HC configuration; panels (b), (d), (f), (h) correspond to $f_2$, where both qubits are biased at the NHC configuration. The black solid curves overlaid on the phase response [(e)-(h)] are eigenvalues calculated from \eqref{eqefip} as a function of $f$, tracing the power-dependent activation and suppression of individual Floquet coupling channels and directly mapping the non-monotonic Bessel-function envelope onto the quasienergy dispersions. The horizontal dashed lines in (e) and (g) mark the $\delta$ values at which the corresponding $g_n$ vanish. The numbers above the lower panels [(e)-(h)] label representative Floquet channels, with $n=0$ corresponding to the central band, $n=\pm1$ to the first-order sidebands, and the remaining unlabeled branches corresponding to higher-order sidebands.
}
\label{fig:Fig3_sup}
\end{figure}

\figref{fig:Fig3_sup} shows the numerical simulations corresponding to Figs. 2(e)-(h) of the main text, presenting all four pumping configurations for both operating frequencies, $f_1$ and $f_2$, as a function of $\delta=\Delta_i/\Delta_0$, and $i=m$, $d$ for the mirror and distant qubits, respectively. $\Delta_0$ is the modulation amplitude corresponding to the reference pump power used in the experiment. At $f_1$, the progressive activation of higher-order Floquet sidebands with increasing $\delta$ and the non-monotonic Bessel-function envelope of the splitting amplitudes are clearly reproduced for both $Q_M$-pumped and $Q_D$-pumped cases. The asymmetry between the two pumping configurations at $f_1$ [\figand{fig:Fig3_sup}{a}{c}] remains visible: the $Q_D$-pumped panels exhibit a more clearly resolved Bessel envelope since the spectral evolution is observable only through the HC channel $g_n$, without $Q_D$'s own sideband background. At $f_2$, increasing $P_{pump}$ modulates the collective linewidth through the correlated decay channel without generating coherent splittings. The black curves overlaid on the phase response panels indicate the Floquet sideband resonance positions analytically predicted by \eqref{eqefip}, showing good agreement with the numerical results across both pumping configurations and both operating frequencies. Meanwhile, the horizontal dashed lines mark the $\delta$ values at which the corresponding $g_n$ vanish, indicating the zeros of the associated Floquet coupling channels.

\begin{figure}[h]
\includegraphics[width=\linewidth]{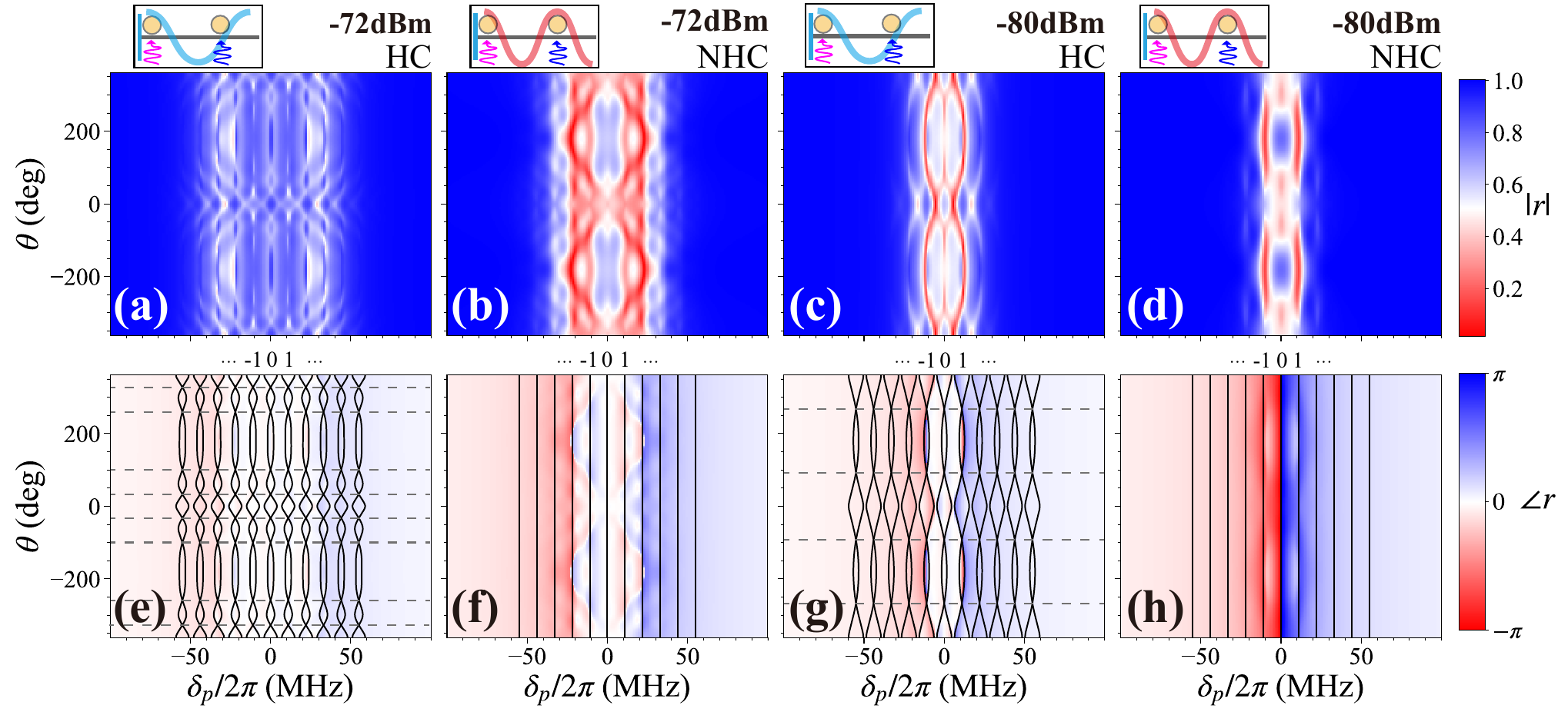}
\caption{
Numerical simulations of the Floquet-engineered qubit-qubit coupling as a function of $\theta$, corresponding to Fig. 3 in the main text. The coupling mechanism is illustrated in Fig. 1(e) of the main text. (a)-(d) Magnitude of reflection coefficient $\left|r\right|$ and (e)-(h) phase response $\angle r$ as a function of $\delta_p$ and $\theta$ at fixed $\nu/2\pi=\unit[11]{MHz}$. Flux pumps are applied simultaneously to both $Q_M$ and $Q_D$, as illustrated by the pump-condition schematics above each panel. Panels (a), (e) correspond to $f_1$ at $P_{pump}\approx\unit[-72]{dBm}$ and panels (c), (g) correspond to $f_1$ at $P_{pump}\approx\unit[-80]{dBm}$; panels (b), (f) correspond to $f_2$ at $P_{pump}\approx\unit[-72]{dBm}$ and panels (d), (h) correspond to $f_2$ at $P_{pump}\approx\unit[-80]{dBm}$. The black solid curves overlaid on the phase response [(e)-(h)] are eigenvalues calculated from \eqref{eqefi} for the two-qubit co-driven case, tracing $\theta$-periodic modulation of the Floquet quasienergy dispersion through the argument $2f\sin(\theta/2)$ of the Bessel function. The horizontal dashed lines in (e) and (g) mark the $\theta$ values at which the corresponding $g_n$ vanish. The numbers above the lower panels [(e)-(h)] label representative phase-dependent Floquet channels, including the central band $n=0$ and first-order sidebands $n=\pm1$; additional branches arise from higher-order sidebands.
}
\label{fig:Fig4_sup}
\end{figure}

\figref{fig:Fig4_sup} shows numerical simulations corresponding to Fig. 3 of the main text, presenting all four combinations of pump power and operating frequency as a function of relative pump phase $\theta$. The simulations capture the main observed trends in the experimental data, including the $\theta$-periodic modulation of the coherent exchange splitting at $f_1$ and the collective linewidth modulation at $f_2$. At $f_1$, the richer multi-zero interference pattern at higher pump power compared to lower pump power reflects the activation of more Bessel-function zeros within the swept $\theta$ range, consistent with the phase-controlled Bessel argument $2f\sin{(\theta/2)}$ in \eqref{eqefi}. The black curves overlaid on the phase response panels indicate the Floquet quasienergy branches calculated from \eqref{eqefi}, showing good agreement with the numerical results across all four configurations. Meanwhile, the horizontal dashed lines mark the $\theta$ values at which corresponding $g_n$ vanish, providing a direct guide to the suppression points of the associated coupling channels. 
\clearpage

\bibliography{SI}